\documentclass[12pt,preprint]{aastex}
\usepackage{graphicx}
\usepackage{subfigure}
\slugcomment{Accepted by ApJ}
\newcommand{\detg}{{\sqrt{-g}}}
\newcommand{\del}{{\partial}}
\newcommand{\KS}{{\rm KS}}
\newcommand{\BL}{{\rm BL}}

\newcommand{\msun}{{\rm M_{\odot}}}
\newcommand{\lsun}{{\rm L_{\odot}}}

\newcommand{\dF}{{^{^*}\!\!F}}

\newcommand{\bnabla}{{\bf \nabla}}
\newcommand{\bB}{{\bf B}}

\newcommand{\eps}{\epsilon}
\newcommand{\order}{{\mathcal{O}}}

\newcommand{\eflux}{\dot{E}}
\newcommand{\efluxem}{{\dot{E}}^{(EM)}}
\newcommand{\efluxma}{{\dot{E}}^{(MA)}}
\newcommand{\lfluxem}{{\dot{L}}^{(EM)}}
\newcommand{\lfluxma}{{\dot{L}}^{(MA)}}
\newcommand{\lflux}{\dot{L}}
\newcommand{\mflux}{{\dot{M}}_0}
\newcommand{\dlflux}{{F_L}}
\newcommand{\dmflux}{{F_M}}
\newcommand{\deflux}{{F_E}}
\newcommand{\defluxem}{{F}^{(EM)}_E}
\newcommand{\defluxma}{{F}^{(MA)}_E}
\newcommand{\dlfluxem}{{F}^{(EM)}_L}

\newcommand{\Trt}{{{T^{r}_t}}}

\newcommand{\IEDEN}{u}
\received{December 17, 2003}

\revised{April 22, 2004}

\accepted{April 26, 2004}

\shortauthors{McKinney and Gammie}
\shorttitle{Black Hole Luminosity}

\begin{document}

\title{A Measurement of the Electromagnetic Luminosity of a Kerr Black Hole}

\author{Jonathan C. McKinney$^{1,2}$ and Charles F. Gammie$^{1,2,3}$}

\altaffiltext{1}{Center for Theoretical
Astrophysics, University of Illinois at Urbana-Champaign,
Loomis Laboratory, 1110 W.Green St. Urbana,IL 61801}

\altaffiltext{2}{Department of Physics}

\altaffiltext{3}{Department of Astronomy}

\email{jcmcknny@uiuc.edu, gammie@uiuc.edu}

\begin{abstract}

Some active galactic nuclei, microquasars, and gamma ray bursts
may be powered by the electromagnetic braking of a rapidly
rotating black hole. We investigate this possibility via
axisymmetric numerical simulations of a black hole surrounded by a
magnetized plasma.  The plasma is described by the equations of
general relativistic magnetohydrodynamics, and the effects of
radiation are neglected. The evolution is followed for $2000 G
M/c^3$, and the computational domain extends from inside the event
horizon to typically $40 G M/c^2$.  We compare our results to two
analytic steady state models, including the force-free
magnetosphere of Blandford \& Znajek.  Along the way we present a
self-contained rederivation of the Blandford-Znajek model in
Kerr-Schild (horizon penetrating) coordinates.  We find that (1)
low density polar regions of the numerical models agree well with
the Blandford-Znajek model; (2) many of our models have an outward
Poynting flux on the horizon in the Kerr-Schild frame; (3) none of
our models have a net outward energy flux on the horizon; and (4)
one of our models, in which the initial disk has net magnetic
flux, shows a net outward angular momentum flux on the horizon. We
conclude with a discussion of the limitations of our model,
astrophysical implications, and problems to be addressed by future
numerical experiments.

\end{abstract}


\maketitle

\section{Introduction}

A black hole of mass $M$ and angular momentum $J = a G M/c$, $0 \le a/M
< 1$ has a free energy associated with its angular momentum (or
``spin'').  This energy can, in principle, be tapped by manipulating
particle orbits so that negative energy particles are accreted
\citep{pen69}.  Spin energy can also be tapped by superradiant
scattering of vacuum electromagnetic waves \citep{pt72}, gravity waves
\citep{hh72,tp74}, or magnetohydrodynamic (MHD) waves \citep{u97}.  It
can also be tapped through the action of force-free electromagnetic
fields \citep{bz77}.

The Blandford-Znajek (BZ) effect-- broadly used here to mean the
extraction of energy from rotating holes via a magnetized plasma--
appears to be the most astrophysically plausible exploitation of
black hole spin energy.  Relativistic jets in active galactic
nuclei, galactic microquasars, and gamma-ray bursts (GRBs) may
well be powered by the BZ effect. Despite some hints (e.g.
\citealt{wilms2001}, \citealt{miller2002}, \citealt{mt03}) and the
general consistency of this idea with the data, however, there is
no {\it direct} observational evidence for black hole energy
extraction.  In this paper we take an experimental approach and
study the BZ effect through direct numerical simulation of a
magnetized plasma accreting onto a black hole.

The energy stored in black hole spin is potentially large.  If
$M_{irr}$ is the ``irreducible mass'' of the black hole where, in
units such that $G = c = 1$,
\begin{equation}
M_{irr}^2 =  {1\over{2}} M r_+ ,
\end{equation}
and $r_+ = M (1 + \sqrt{1 - (a/M)^2})$ is the horizon radius, then
the free energy is
\begin{equation} E_{spin} = M - M_{irr} < 5.3 \times
10^{61} \left({M \over{10^8 \msun}}\right) {\rm erg}.
\end{equation}
or $\approx 30\%$ of the gravitational mass of a maximally rotating
hole.  This corresponds to a luminosity of $\lesssim 4 \times 10^{10}
(M/10^8\msun) \lsun $ if released over a Hubble time.

Estimates suggest that black hole accretion is surprisingly
efficient, in the sense that the ratio of quasar radiative energy
density to supermassive black hole mass density is $\sim 0.2$
\citep{yt02,erz02}. During the accretion process some mass-energy
is radiated away and the rest is incorporated into the black hole.
Through electromagnetic spindown this energy gets a second chance
to escape.  A combination of efficient thin disk accretion (in
which all radiation is somehow permitted to escape) followed by
the Penrose process can in principle extract up to $(1 -
1/\sqrt{6}) c^2 = 0.59 c^2$ per gram of accreted rest-mass.  In
practice, of course, much less energy is likely to be available.
One goal of our investigation is to discover how much less. Part
of the answer may lie with the calculations already described in
\cite{gsm2004}: if black hole spins are limited by the equilibrium
value found there ($a/M \approx 0.92$) then the nominal thin disk
efficiency of the accretion phase is about $\approx 17\%$, much
less than the 42\% expected at $a/M=1$.

In this paper we consider the self-consistent evolution of a
weakly magnetized torus surrounding a rotating black hole.  The
evolution is carried out numerically in the axisymmetric ideal MHD
approximation.  As the evolution progresses the computational
domain develops matter dominated regions near the equator and
electromagnetic field dominated regions near the poles.  To fix
expectations for the structure of these regions we review two
analytic models for the interaction of a magnetized plasma with a
black hole in \S~\ref{analyticmodels}.  Along the way we develop
the relevant notation and coordinate systems.  In
\S~\ref{numericalmodels} we describe our numerical model and give
a summary of numerical results for a high resolution fiducial
model. In \S~\ref{parameterstudy} we consider the dependence of
our results on model parameters. A discussion and summary may be
found in \S~\ref{bzdiscussion}. From here on we adopt units such
that $G M = c = 1$. Table~\ref{tbl1} gives a list of commonly used
symbols.

\begin{deluxetable}{lll}
\tablewidth{0pt}

\tablecaption{Commonly used symbols \label{tbl1}}

\tablehead{

\colhead{Symbol}

& \colhead{Fiducial Value}

& \colhead{Description}

}

\startdata
& \underline{Model Parameters} & \\
 $a$ & $0.938$ & black hole spin ($J/M^2$) \\
 $r_+$ &  $1.347$ & radius of the event horizon ($r_+=1+\sqrt{1-a^2}$) \\
 $r_{isco}$ &  $2.044$ & radius of the ISCO (innermost stable circular orbit) \\
 $r_{edge}$ &  $6$ & radius of inner edge of torus\\
 $r_{max}$ &  $12$ & radius of the pressure maximum \\
 $\Omega_H$ &  $\approx 0.3477$ & spin frequency of zero angular momentum observer at $r_+$ \\
 $R_{in}$ &  $0.98 r_+$ & inner radial grid location \\
 $R_{out}$ &  $40$ & outer radial grid location \\
 $\beta$ & $100$ & ratio of gas to magnetic pressure (initially $\frac{p_{gas,max}}{p_{mag,max}}$) \\
 $\gamma$ & $4/3$ & $p_{gas}=(\gamma-1)u$ \\
 & &\\
& \underline{Diagnostics} & \\
 $\mflux$ & see sections~\ref{goveqns}     \& \ref{numericalmodels} & rest-mass flux into the black hole \\

 $\eflux$ & see sections~\ref{goveqns}     \& \ref{numericalmodels} & energy flux into the black hole \\
 $\efluxem$ & see sections~\ref{goveqns}   \& \ref{numericalmodels} & electromagnetic energy flux \\
 $\efluxma$ & see sections~\ref{goveqns}   \& \ref{numericalmodels} & matter energy flux \\

 $\lflux$ & see sections~\ref{goveqns}     \& \ref{numericalmodels} & angular momentum flux into the black hole \\
 $\lfluxem$ & see sections~\ref{goveqns}   \& \ref{numericalmodels} & electromagnetic angular momentum flux \\
 $\lfluxma$ & see sections~\ref{goveqns}   \& \ref{numericalmodels} & matter angular momentum flux \\
 $\tilde{L}$ & see sections~\ref{fiducial}  \& \ref{parameterstudy} & $\tilde{L}=\efluxem/(-\eps\mflux)$ ; $\eps = 1 - \eflux/\mflux$\\
 & &\\
& \underline{Variables} & \\
 $b^2/2$ &  see section~\ref{inflowcomp} & electromagnetic energy density in the fluid frame\\
 $B^r$,$B^\theta$,$B^\phi$ & see section~\ref{goveqns} & magnetic field
 components. $B^i = \dF^{it}$ \\
 $A_\phi$ & see section~\ref{numericalmodels} & azimuthal component of electromagnetic
 vector potential \\
 $\tilde{v}^r$ & see section~\ref{fiducial}   & asymptotic radial velocity (i.e. $v^r$ at $r=\infty$) \\
 $\omega$ &  see sections~\ref{bzcomp} \& ~\ref{inflowcomp} & spin frequency of electromagnetic field \\
 $\Omega$ &  see section~\ref{inflowcomp} & spin frequency of fluid ($\Omega=u^\phi/u^t$)\\

\enddata
\end{deluxetable}

\section{Review of Analytic Models}\label{analyticmodels}

In this section we review two quasi-analytic, steady state models
for the interaction of a black hole with the surrounding plasma.
 The purpose of this review is to introduce our coordinate system
and notation and to describe the models in a form suitable for
later comparison with numerical results.  Along the way, we give a
self-contained derivation of the BZ effect in Kerr-Schild (horizon
penetrating) coordinates.  To the extent that the analytic and
numerical models agree, the comparison also builds confidence in
the numerical models.

\subsection{Coordinates}

Before proceeding it is useful to define three coordinate bases for the
Kerr metric.

{\it Boyer-Lindquist (BL) coordinates.}  These are the most familiar
coordinates for the Kerr metric.  In BL coordinates $t,r,\theta,\phi$
\begin{equation}
ds^2 = - \left( 1 - \frac{2\,r}{\Sigma}\right)\,dt^2 +
  \frac{\Sigma}{\Delta}\,dr^2 +
  \Sigma\,d\theta^2 + \frac{A\,{\sin^2
  {\theta}}}{\Sigma}\,d\phi^2 - \frac{4\,a\,r\,{\sin^2
  {\theta}}}{\Sigma}\,d\phi\,dt
\end{equation}
where $\Sigma\equiv r^2+a^2\cos^2{\theta}$, $\Delta\equiv r^2 - 2 r +
a^2$ and $A\equiv (r^2+a^2)^2- a^2\Delta\sin^2\theta$.  The determinant
of the metric $g \equiv {\rm Det}(g_{\mu\nu}) = -\Sigma^2 \sin^2
\theta$.  In BL coordinates the metric is singular on the event horizon
at $r = r_+$ where $\Delta = 0$.

{\it Kerr-Schild (KS) coordinates.}  The Kerr-Schild coordinates
$t,r,\theta,\phi$ are regular on the horizon. They are closely
related to BL coordinates: $r[{\rm KS}] = r[{\rm BL}]$ and $\theta
[{\rm KS}] = \theta [{\rm BL}]$.  The line element is
$$
ds^2
= -\left( 1 - \frac{2\,r}{{\Sigma}} \right)\,dt^2
+ \left(\frac{4\,r}{{\Sigma}}\right)\,dr\,dt
+ \left( 1 + \frac{2\,r}{\Sigma} \right)\,dr^2
+ {\Sigma}\,d\theta^2
$$
$$
 + {\sin^2 {\theta}}\,\left( {\Sigma}
 + a^2\,\left( 1 + \frac{2\,r}{{\Sigma}} \right)
 \,{\sin^2 {\theta}} \right)\,d\phi^2
$$
\begin{equation}\label{ksmetric}
- \left(\frac{4\,a\,r\,{\sin^2 {\theta}}}{{\Sigma}}\right)\,d\phi\,dt
- 2\,a\,\left( 1 + \frac{2\,r}{{\Sigma}}\right)\,{\sin^2
  {\theta}}\,d\phi\,dr,
\end{equation}
and $g = -\Sigma^2 \sin^2 \theta$.

The transformation matrix from BL to KS is
\begin{equation}
{\del t[\KS]\over{\del r[\BL]}} = {2 r\over{\Delta}},
\end{equation}
and
\begin{equation}
{\del \phi[\KS]\over{\del r[\BL]}} = {a\over{\Delta}};
\end{equation}
all other off-diagonal components are $0$ and all diagonal components
are $1$.  The inverse transformation matrix is identical, with the signs
of the off-diagonal components reversed.

{\it Modified Kerr-Schild (MKS) coordinates.} Our numerical integrations
are carried out in a modified KS coordinates $x_0, x_1, x_2, x_3$, where
$x_0 = t[{\rm KS}]$, $x_3 = \phi[{\rm KS}]$, and
\begin{equation}\label{radius}
r = e^{x_1},
\end{equation}
\begin{equation}\label{theta}
\theta = \pi x_2 + \frac{1}{2} (1 - h) \sin(2\pi x_2).
\end{equation}
Here $h$ is an adjustable parameter that can be used to concentrate grid
zones toward the equator as $h$ is decreased from $1$ to $0$.  The
transformation matrix from KS to MKS is diagonal and trivially
constructed from the explicit expressions for $r$ and $\theta$ in
equations~\ref{radius} and~\ref{theta}.

\subsection{Governing Equations}\label{goveqns}

For a magnetized plasma the equations of motion are
\begin{equation}\label{EOM}
{T^{\mu\nu}}_{;\nu}
= \left(T^{\mu\nu}_{\rm MA} + T^{\mu\nu}_{\rm EM}\right)_{;\nu}
= 0.
\end{equation}
where $T^{\mu\nu}$ is the stress-energy tensor, which can be split into
a matter (MA) and electromagnetic (EM) part.  In the fluid approximation
\begin{equation}
T^{\mu\nu}_{\rm MA} = (\rho_0 + \IEDEN + p) u^\mu u^\nu + p
g^{\mu\nu},
\end{equation}
where $\rho_0 \equiv$ rest-mass density, $\IEDEN \equiv$ internal
energy, $p \equiv$ pressure, $u^\mu$ is the fluid four-velocity, and
we assume throughout an ideal gas equation of state
\begin{equation}
p = (\gamma - 1) \IEDEN.
\end{equation}
 In terms of $F^{\mu\nu}$, the Faraday (or
electromagnetic field) tensor,
\begin{equation}\label{tmunuem}
T^{\mu\nu}_{\rm EM} =
F^{\mu\gamma}{F^{\nu}}_{\gamma} -\frac{1}{4}g^{\mu\nu}
F^{\alpha\beta}F_{\alpha\beta},
\end{equation}
where we have absorbed a factor of $\sqrt{4\pi}$ into the definition
of $F^{\mu\nu}$.  We assume that particle number is conserved:
\begin{equation}
(\rho_0 u^\mu)_{;\mu} = 0.
\end{equation}
The evolution of the electromagnetic field is given by the space
components of the source-free Maxwell equations
\begin{equation}
{\dF^{\mu\nu}}_{;\nu} = 0,
\end{equation}
where $\dF$ is the dual of the Faraday, and the time component
gives the no-monopoles constraint.  The inhomogeneous Maxwell
equations
\begin{equation}
J^\mu = {F^{\mu\nu}}_{;\nu}
\end{equation}
define the current density $J^\mu$ but are otherwise not required here.
We adopt the ideal MHD approximation, where
\begin{equation}
u_\mu F^{\mu\nu} = 0,
\end{equation}
which implies that the electric field vanishes in the rest frame of the
fluid.

In our numerical models the fundamental (or ``primitive'') variables
that describe the state of the plasma are $\rho_0, \IEDEN, B^i \equiv
\dF^{it}$, plus three variables which describe the motion of the plasma.
In \cite{gmt2003} we used the plasma three-velocity. Here we use
\begin{equation}
\tilde{u}^i \equiv u^i + {\gamma \beta^i\over{\alpha}},
\end{equation}
where $\gamma \equiv \sqrt{1+q^2}$, $q^2 \equiv g_{ij} \tilde{u}^i
\tilde{u}^j$, $\beta^i \equiv g^{ti} \alpha^2$ is the shift, and
$\alpha^2 = -1/g^{tt}$ is the lapse.  We made this change to improve
numerical stability.  Because the three velocity components have a
finite range, truncation error can move the plasma velocity outside the
light cone.  The variables $\tilde{u}^i$ have the important property
that they range over $-\infty$ to $\infty$, and this makes it impossible
for the plasma to step outside the light cone.

To write the electromagnetic quantities in terms of the primitive
variables, define the four-vector $b^\mu$ with $b^t \equiv
g_{i\mu} B^i u^\mu$ and $b^i \equiv (B^i + u^i b^t)/u^t$. With
some manipulation one finds
\begin{equation}
T^{\mu\nu}_{\rm EM} =
b^2 u^\mu u^\nu + {b^2\over{2}} g^{\mu\nu} - b^\mu b^\nu,
\end{equation}
and
\begin{equation}
\dF^{\mu\nu} = b^\mu u^\nu - b^\nu u^\mu.
\end{equation}
The no-monopoles constraint becomes
\begin{equation}
(\detg B^i)_{,i} = 0.
\end{equation}
A more complete account of the relativistic MHD equations can be
found in \cite{gmt2003} or \cite{anile}.

\subsection{Blandford-Znajek Model}\label{bzanalytic}

BZ studied a rotating black hole surrounded by a stationary,
axisymmetric, force-free, magnetized plasma.  They obtain an
expression for the energy flux through the event horizon and,
given a solution for the field geometry when $a = 0$, find a
perturbative solution when $a \ll 1$.  Here we present a
self-contained rederivation, which will be compared to numerical
models in section~\ref{bzcomp}.  Those not interested in the
derivation may find a summary set of equations in \ref{bzsummary}.
A comparison of the analytic BZ model to our numerical models can
be found in section~\ref{bzcomp}.

We follow an approach that differs slightly from BZ.  We solve
${T^{\mu\nu}}_{;\nu}=0$ directly rather than using
$J_{\mu}F^{\mu\nu}=0$, which is equivalent in the force-free
approximation.  Also, because our solution is developed in KS
coordinates, which are regular on the horizon, we obtain the BZ
solution by applying a regularity condition on the horizon and at
large radius, rather than the physically equivalent approach of
applying a regularity condition on the horizon in the Carter
tetrad \citep{z77} and then applying the result as a boundary
condition in BL coordinates.  Finally, if we assume separability
of the solution then we do not need to require that the solution
match the flat-space force-free solution of \citet{michel1973}.

\subsubsection{Derivation in KS coordinates}

Over the poles of the black hole it is reasonable to expect that the
density is low, but the field strength is comparable to that at the
equator.  In the limit that
\begin{equation}\label{HIGHB}
b^2 \gg \rho_0 + \IEDEN + p,
\end{equation}
where $b^2$ is the field strength in the fluid frame, one may
assume that the matter contribution to the stress energy tensor
can be ignored and
\begin{equation}\label{ffapprox}
T^{\mu\nu} \approx T^{\mu\nu}_{\rm EM}.
\end{equation}
This is the force-free limit.

The ideal MHD condition $u^\mu F_{\mu\nu} = 0$ implies that the
electric field vanishes in the rest frame of the fluid.  Therefore
the invariant ${\bf E}\cdot {\bf B} = 0$, or in covariant form
$\dF^{\mu\nu} F_{\mu\nu} = 0$. The electromagnetic field is then
said to be degenerate.

In the force-free limit the governing equations are then
\begin{equation}
T^{\mu\nu}_{\rm EM; \nu} = 0
\end{equation}
and
\begin{equation}
{\dF^{\mu\nu}}_{;\nu} = 0.
\end{equation}
As BZ point out, the same basic set of equations can be derived without
assuming that the plasma obeys the fluid equations.

We now specialize to KS coordinates and write down the Faraday tensor in
terms of a vector potential $A_\mu$, $F_{\mu\nu} = A_{\nu,\mu} -
A_{\mu,\nu}$.  We assume that the field is axisymmetric ($\del_\phi
\rightarrow 0$) and stationary ($\del_t \rightarrow 0$).  Evaluating the
condition $\dF^{\mu\nu} F_{\mu\nu} = 0$, one finds
\begin{equation}
A_{\phi,\theta} A_{t,r} - A_{t,\theta} A_{\phi,r} = 0.
\end{equation}
It follows that one may write
\begin{equation}\label{omega}
{A_{t,\theta}\over{A_{\phi,\theta}}} =
{A_{t,r}\over{A_{\phi,r}}} \equiv -\omega(r,\theta)
\end{equation}
where $\omega(r,\theta)$ is an as-yet-unspecified function.  It is
usually interpreted as the ``rotation frequency'' of the
electromagnetic field (this is Ferraro's law of isorotation; see
e.g. \citealt{fkr}, \S 9.7 in a nonrelativistic context).  This
yields $F_{\mu\nu}$ in terms of the free functions $\omega,
A_\phi$, and $B^\phi$, the toroidal magnetic field:
\begin{equation}\label{startfmunu}
F_{tr} = -F_{rt} = \omega A_{\phi,r}
\end{equation}
\begin{equation}
F_{t\theta} = -F_{\theta t} = \omega A_{\phi,\theta}
\end{equation}
\begin{equation}
F_{r\theta} = -F_{\theta r} = \detg B^\phi
\end{equation}
\begin{equation}
F_{r\phi} = -F_{\phi r} = A_{\phi,r}
\end{equation}
\begin{equation}\label{endfmunu}
F_{\theta\phi} = -F_{\phi \theta} = A_{\phi,\theta}
\end{equation}
with all other components zero.  Written in this form, the
electromagnetic field automatically satisfies the source-free Maxwell
equations.  Notice that $A_{\phi,\theta}=\detg B^r$ and
$A_{\phi,r}=-\detg B^{\theta}$.

We want to evaluate the radial energy flux
\begin{equation}\label{efluxexpression}
\eflux \equiv 2\pi\int_0^{\pi} \, d\theta \,
\detg \deflux
\end{equation}
where $\deflux\equiv -\Trt$.  This can be subdivided into a matter
$\defluxma$ and electromagnetic $\defluxem$ part, although in the
force-free limit the matter part vanishes.  Similar expressions
can be written for the angular momentum flux $\lflux$ and angular
momentum flux density $\dlflux$, and for the mass flux $\mflux$
and mass flux density $\dmflux$.  In the limit of a steady flow
these conserved quantities correspond to the radial flux measured
by a stationary observer at large distance from the black hole.

Using the definition of the electromagnetic stress-energy tensor
(\ref{tmunuem}) and the relations (\ref{startfmunu})-(\ref{endfmunu}),
it is a straightforward exercise to evaluate
\begin{equation}
\defluxem = -2 (B^r)^2 \omega r (\omega - {a\over{2 r}}) \sin^2{\theta}
- B^r B^\phi \omega \Delta \sin^2{\theta}.
\end{equation}
The radial angular momentum flux density is $\dlfluxem =
\defluxem/\omega$.  One can verify by direct transformation that
$\deflux[{\rm KS}] = \deflux[{\rm BL}]$ and $\dlflux[\KS] =
\dlflux[\BL]$.  On the horizon $r = r_+ = 1 + \sqrt{1 - a^2}$ and
$\Delta = 0$, so the horizon energy flux is
\begin{equation}\label{HORFLUX}
\defluxem|_{r = r_+} = 2 (B^r)^2 \omega r_+ (\Omega_H - \omega) \sin^2{\theta}
\end{equation}
where $\Omega_H \equiv a/(2 r_+)$ is the rotation frequency of the black
hole (see MTW \S 33.4).  This result, which is identical to BZ's result,
implies that if $0 < \omega < \Omega_H$ and $(B^r)^2 > 0$ then there is
an outward directed energy flux at the horizon.  Because the flux was
evaluated in KS coordinates the horizon did not require special treatment
as in \cite{z77}.

To finish evaluating $\efluxem$ we need to find $A_\phi$, $\omega$, and
$B^\phi$.  This requires solving the equations of motion (\ref{EOM}).
They can be evaluated directly or in the reduced form $J^\mu F_{\mu\nu}
= 0$ (as in BZ), in which case one must also evaluate the currents using
Maxwell's equations.  In either form this is a difficult, nonlinear
problem which probably cannot be solved in any general way.

To make progress, BZ find solutions to the equations of motion when $a =
0$, then perturb them by allowing the black hole to spin slowly with $a
\ll 1$.  If we assume that the initial field has $\omega = B^\phi = 0$,
then we may expand the vector potential
\begin{equation}
A_\phi = A_\phi^{(0)}(r,\theta) + a^2 A_\phi^{(2)}(r,\theta) + \order(a^4),
\end{equation}
where $A_\phi^{(1)} = 0$ by symmetry ($A_\phi$ should be even in $a$).
The field rotation frequency vanishes in the unperturbed solution, and
$\omega^{(2)} = 0$ because $\omega$ should be odd in $a$, so
\begin{equation}
\omega = a \omega^{(1)}(r,\theta) + \order(a^3)
\end{equation}
and similarly for the toroidal field
\begin{equation}
B^{\phi} = a B^{\phi (1)}(r,\theta) + \order(a^3).
\end{equation}
We are now in a position to find the free functions $A_\phi^{(2)},
\omega^{(1)},$ and $B^{\phi (1)}$, given an initial field $A_\phi^{(0)}$
that satisfies the basic equations when $a = 0$.

BZ consider two forms for $A_\phi^{(0)}$: a monopole field and a
paraboloidal field.  Here we review only the (possibly split)
monopole, where $A_\phi^{(0)} = -C \cos{\theta}$ and $C$ is an
arbitrary constant. One may obtain the perturbed solution by
making the following sequence of deductions.

(1) The $t$ and $\phi$ components of equation (\ref{EOM}), expanded to
lowest nontrivial order in $a$, require that $\dlfluxem$ and $\defluxem$
be independent of radius.  Therefore they are functions of $\theta$
alone.  Since
\begin{equation}
\defluxem = a \omega^{(1)} \dlfluxem,
\end{equation}
we conclude that $\omega^{(1)}$ is a function of $\theta$ alone.

(2) The $r$ component of equation (\ref{EOM}), together with the
requirement that $B^{\phi (1)}$ be finite on the horizon (all components
of $F_{\mu\nu}$ are well-behaved on the horizon in KS coordinates),
yields a single nontrivial solution:
\begin{equation}
B^{\phi (1)} = - {C\over{4 r^2}} \left( 1 - 4 \omega^{(1)} +
{2\over{r}} \right)
\end{equation}
This solution is well behaved at the horizon and at large radius as long
as $\omega^{(1)}$ is finite on the horizon and grows less rapidly than
$r^2$ at large $r$.

(3) The $\theta$ component of equation (\ref{EOM}), which is the
trans-field force balance equation, can now be reduced to an equation
involving $A_\phi^{(2)}$ and $\omega^{(1)}$.  If we require that
$A_\phi^{(2)} = C f(r) g(\theta)$, then one may deduce that (a)
$\del_\theta \omega^{(1)} = 0$, i.e. $\omega^{(1)} = const.$; (b)
$g(\theta) = \cos\theta \sin^2\theta$.  Then $f(r)$ must satisfy
\begin{equation}\label{bzeq}
f'' +
{2 f'\over{r (r - 2)}} -
{6 f\over{r (r - 2)}} +
\left( {r + 2\over{r^3 (r - 2)}} -
{(\omega^{(1)} - 1/8) (r^2 + 2 r + 4)\over{r (r - 2)}} \right) = 0
\end{equation}
which is equivalent to BZ's equation~(6.7).  This has an exact solution
with two constants of integration.  One of the constants of integration
is set by requiring that the solution be finite on the horizon.  Part of
the solution can be regularized at large $r$ by fixing the other
constant of integration, but the remaining divergence can only be zeroed
by setting $\omega^{(1)} = 1/8$; this is already suggested by the form
of the preceding equation.  For $r > 2$ the regular solution is
\begin{eqnarray}\label{FRSOL}
f(r) = \left( {\rm Li}_2(\frac{2}{r}) -
\ln(1 - \frac{2}{r})\ln\frac{r}{2} \right)
{r^2 (2 r - 3)\over{8}}
+ {1 + 3 r - 6 r^2\over{12}}\ln \frac{r}{2} +
 {11\over{72}} + {1\over{3 r}} + {r\over{2}} - {r^2\over{2}},
\end{eqnarray}
where ${\rm Li}_2$ is the second polylogarithm function:
\begin{equation}
{\rm Li}_2(x) = -\int_0^1\,dt\, {\ln (1 - t x)\over{t}}.
\end{equation}
For $r < 2$ the solution is given by the real part of equation
(\ref{FRSOL}).  In the limit of large $r$
\begin{equation}
f(r) \sim {1\over{4 r}} + \order\left({\ln r\over{r^2}}\right),
\end{equation}
which agrees with BZ.

To sum up, using only the assumption of separability of $A_\phi^{(2)}$
and the regularity of physical quantities in Kerr-Schild coordinates on
the horizon and at infinity, we find
\begin{equation}\label{BZw}
\omega^{(1)} = \frac{1}{8}
\end{equation}
\begin{equation}\label{BZBp}
B^{\phi (1)} = -{C\over{8 r^2}}(1 + {4\over{r}})
\end{equation}
and
\begin{equation}
A_\phi^{(2)} = C f(r) \cos\theta\sin^2\theta.
\end{equation}
with $f(r)$ given by equation (\ref{FRSOL}).  Our solution is
identical to BZ's after transforming to Boyer-Lindquist
coordinates and transforming from our $B^\phi$ to BZ's $B_T$,
although BZ's expression for $f(r)$ contains some unclosed
parentheses.

\subsubsection{BZ Derivation Summary}\label{bzsummary}

In Kerr-Schild coordinates, then, the magnetic field components are
\begin{equation}\label{BZBr}
B^r = {C\over{r^2}} + a^2 \frac{C}{2r^4}\left(-2\cos{\theta} +
r^2(1+3\cos{2\theta}) f(r)\right),
\end{equation}
\begin{equation}\label{BZBh}
B^{\theta} = -a^2 \frac{C}{r^2}\cos{\theta}\sin{\theta} f',
\end{equation}
both accurate through second order in $a$, and
\begin{equation}
B^{\phi} = -a {C\over{8 r^2}}(1 + {4\over{r}}),
\end{equation}
accurate through first order in $a$.  In Boyer-Lindquist coordinates,
\begin{equation}
B^r[\BL] = B^r[\KS],
\end{equation}
\begin{equation}
B^\theta[\BL] = B^\theta[\KS],
\end{equation}
\begin{equation}
B^\phi[\BL] = B^\phi[\KS] - B^r[\KS] {(a - 2 r\omega)\over{\Delta}},
\end{equation}
and BZ's toroidal field
\begin{equation}
B_T = \Delta \sin^2\theta B^\phi[\BL]
\end{equation}
(which is different from BZ's $B_\phi$).

There has been some concern about causality in the application of
the force-free approximation (e.g. \citealt{pun2003}, see also
\citealt{kom02,kom04a}).  The MHD equations are hyperbolic and
causal (as are the equations of force-free electrodynamics).  Below
we show that a numerical evolution of the MHD equations agrees well
with the BZ solution in those regions where $b^2/\rho_0 \gg 1$.
This is either a remarkable coincidence or else the BZ solution is
an accurate representation of the strong-field limit of ideal MHD.

For comparison with computational models, the most relevant
aspects of the BZ theory are that: (1) the field is force-free;
(2) the field rotation frequency $\omega = a/8 + \order{(a^3)}$ in
the monopole geometry case and $\omega = a/8 + \order{(a^3)}$ at
the poles ($\theta = 0, \pi/2$) in the paraboloidal field case
considered by BZ;\footnote{ According to the numerical results of
\cite{kom01} and the argument of \cite{mt82}, $\omega$ adjusts to
$\approx \Omega_H/2$ hole even at large $a$.} (3) if the field
geometry is nearly monopolar and $a$ is small enough that the
expansion to lowest order in $a$ is accurate, then $B^r(\theta)$
is given by equation (\ref{BZBr}); and (4) if the field geometry
is monopolar and $a$ is small, then the energy flux density
$\deflux \propto \sin^2\theta$ on the horizon.  We compare this
analytic BZ model to our numerical models in section~\ref{bzcomp}.

\subsection{Equatorial MHD Inflow}\label{inflowanalytic}

\cite{gam99} considered a stationary, axisymmetric MHD inflow in the
``plunging region'', between the innermost stable circular orbit
(ISCO) and the event horizon.  The flow was assumed to be cold (zero
pressure), nearly equatorial, and to proceed along lines of constant
latitude $\theta$.  The latter assumption ignores the requirement of
cross-field force-balance.  This model is analogous to the
\cite{wd67} model for the solar wind, only turned inside out so that
the wind flows from the disk into the black hole.  The model builds
on earlier work by \cite{tak90}, \cite{phi83}, and \cite{cam86}. The
analytic model derived here will be used to compare to numerical
models in section~\ref{inflowcomp}.

The MHD inflow model is stationary ($\del_t \rightarrow 0$),
axisymmetric ($\del_\phi \rightarrow 0$) and nearly equatorial ($\theta
\approx \pi/2$) so $\del_\theta \rightarrow 0$ by symmetry.  In addition
flow proceeds along lines of constant $\theta$.  As a result the model
is one dimensional with a single independent variable $r$.  The
nontrivial dependent variables are the radial and azimuthal
four-velocity $u^r$ and $u^\phi$, the radial and azimuthal magnetic
field $B^r$ and $B^\phi$, and the rest-mass density $\rho_0$.

With these assumptions the equations of general relativistic MHD
can be integrated completely.  The constancy of energy flux
\begin{equation}
-\detg {T^r}_t = const.,
\end{equation}
and angular momentum flux
\begin{equation}
\detg {T^r}_\phi = const.,
\end{equation}
follow from ${T^\mu}_{\nu;\mu} = 0$.  The source-free Maxwell
equations imply
\begin{equation}
\detg B^r = const.,
\end{equation}
which expresses the constraint $\bnabla \cdot \bB = 0$, and the
relativistic ``isorotation law'',
\begin{equation}
\detg~\dF^{r\phi} = \detg (u^r b^\phi - u^\phi b^r) = const.
\end{equation}
where $b^\mu$ is the magnetic field four-vector (defined above).
Finally, conservation of particle number implies
\begin{equation}
\detg \rho_0 u^r = const.
\end{equation}
These five constants yield five constraints on the five nontrivial
fundamental variables $u^r$, $u^\phi$, $B^r$, $B^\phi$, and
$\rho_0$. Given the constants, and using the constitutive relations
that relate the constants and fundamental variables, one can solve
the resulting set of nonlinear equations for the fundamental
variables at each radius.

The next step is to determine the constants.  The radial magnetic
flux and the rest-mass flux are determined by conditions in the
disk and can be left as free parameters.  The remaining three
degrees of freedom are fixed by imposing boundary conditions.
\cite{gam99} imposed the following conditions: (1) the flow is
regular at the fast point (the flow is automatically regular at
the Alfv\'en point-- see \cite{phi83} for a discussion-- and the
slow point is absent because the flow is cold) ; and (2,3) the
four-velocity components $u^r$ and $u^\phi$ match onto a cold disk
at the ISCO.

Energy can be extracted from the black hole if the Alfv\'en point lies
inside the ergosphere \citep{tak90}.  \cite{gam99} calculated $\eflux$
and $\lflux$ as a function of $a$ and $B^r$ and showed that for even
modest magnetic field strength these were modified from the values
anticipated in classical thin disk theory.  The implications of these
modified fluxes for the structure-- particularly the surface
brightness-- of a thin disk were explored by \cite{ak00}.

For comparison with numerical models, the key predictions of the
inflow model are: (1) the constancy of the conserved quantities
with radius; (2) matching of the flow velocity to circular orbits
at the ISCO; (3) modification of the angular momentum and energy
fluxes from their thin disk values; and (4) the run of all the
fluid variables with radius in the plunging region.

\section{Numerical Experiments}\label{numericalmodels}

All our experiments evolve a weakly magnetized torus around a Kerr
black hole in axisymmetry.  The focus of our numerical
investigation is to study a high resolution model
(\ref{fiducial}), compare with the BZ model (\ref{bzcomp}), and
compare to the Gammie inflow model (\ref{inflowcomp}). In
\S~\ref{parameterstudy} we investigate how various parameters
affect the results.

The initial conditions consist of an equilibrium torus
(\citealt{fm76} ; \citealt{ajs78}) which is a ``donut'' of plasma
with a black hole at the center.  The donut is supported against
gravity by centrifugal and pressure forces, and is embedded in a
vacuum. We consider a particular instance of the \cite{fm76}
solutions, which are defined by the condition $u^t u_\phi =
const.$ We normalize the peak density $\rho_{0,max}$ to $1$ and
fix the inner edge of the torus at $r_{edge} = 6$.  We also set
$\gamma = 4/3$.\footnote{We have run a limited number of
$\gamma=5/3$ models and find results essentially identical to
those discussed below.} Absent a magnetic field, the initial torus
is a stable equilibrium.\footnote{In axisymmetry.  The torus is
unstable to global nonaxisymmetric modes \citep{pp83}.}

Into the initial torus we introduce a purely poloidal magnetic
field. The field can be described using a vector potential with a
single nonzero component $A_\phi \propto {\rm
MAX}(\rho_0/\rho_{0,max} - 0.2, 0)$ The field is therefore
restricted to regions with $\rho_0/\rho_{0,max} > 0.2$.  The field
is normalized so that the minimum ratio of gas to magnetic
pressure is $100$.  The equilibrium is therefore only weakly
perturbed by the magnetic field.  It is, however, no longer stable
\citep{bh91,gam04}.

Our numerical scheme is HARM \citep{gmt2003}, a conservative,
shock-capturing scheme for evolving the equations of general
relativistic MHD.  HARM uses constrained transport to maintain a
divergence-free magnetic field \citep{eh88,toth00}. The inversion of
conserved quantities to primitive variables is performed by solving a
single non-linear equation \citep{dzb} or by a slower but more robust
multi-dimensional Newton-Raphson method. Unless otherwise stated we use
modified Kerr-Schild (MKS) coordinates with $h=0.3$. The computational
domain is axisymmetric, with a grid that typically extends from $r_{in}
= 0.98 r_+$ to $r_{out} = 40$, and from $\theta = 0$ to $\theta =
\pi/2$.

HARM is unable to evolve a vacuum, so we are forced to introduce
``floors'' on the density and internal energy.  When the density
or internal energy drop below these values they are immediately
reset. This sacrifices exact conservation of energy, particle
number, and angular momentum, although it is reasonable to assume
that when the floors are small enough the true solution is
recovered. The floors are position dependent, with $\rho_{0,min} =
10^{-4} r^{-3/2}$ and $\IEDEN_{min} = 10^{-6} r^{-5/2}$.  We
discuss the effect of varying the floor in section~\ref{numparam}.

At the outer boundary we use an ``outflow'' boundary condition.  This
means we project all primitive variables into the ghost zones while
forbidding inflow.  The inner boundary condition is identical except
that, because the boundary is inside the event horizon, we never need to
worry about backflow into the computational domain. At the poles we use
a reflection boundary condition where we impose appropriate symmetries
for each variable across the axis.

\subsection{Fiducial Model}\label{fiducial}

First consider the evolution of a high resolution fiducial model
with $a = 0.938$.  This is close to the spin equilibrium value
(where $d (a/M)/dt = 0$) found by \cite{gsm2004} for a series of
similar Fishbone-Moncrief tori.

The fiducial model has $u^t u_\phi = 4.281$, the pressure maximum
is located at $r_{max} = 12$, the inner edge at
$(r,\theta)=(6,\pi/2)$, and the outer edge at $(r,\theta) = (42,
\pi/2)$.  The orbital period at the pressure maximum $2 \pi (a +
r_{max}^{3/2}) \simeq 267$, as measured by an observer at
infinity.

The numerical resolution of the fiducial model is $456^2$.  The
zones are equally spaced in modified Kerr-Schild coordinates $x_1$
and $x_2$, with coordinate parameters $h=0.3$. Small
perturbations are introduced in the velocity field, and the model
is run for $\Delta t = 2000$, or about $7.6$ orbital periods at
the pressure maximum.

The initial state is Balbus-Hawley unstable.  The inner edge of
the disk quickly makes a transition to turbulence.  Transport of
angular momentum by the magnetic field causes material to plunge
from the inner edge of the disk into the black hole.  The
turbulent region gradually expands outward to involve the entire
disk.  The disk relaxes toward a ``Keplerian'' velocity profile,
meaning that the orbital frequency along the equator is close to
the circular orbit frequency.  The disk enters a long,
quasi-steady phase in which the accretion rates of rest-mass,
angular momentum, and energy onto the black hole fluctuate around
a well-defined mean.

\begin{figure}
\centering
\includegraphics[width=6.49in,clip]{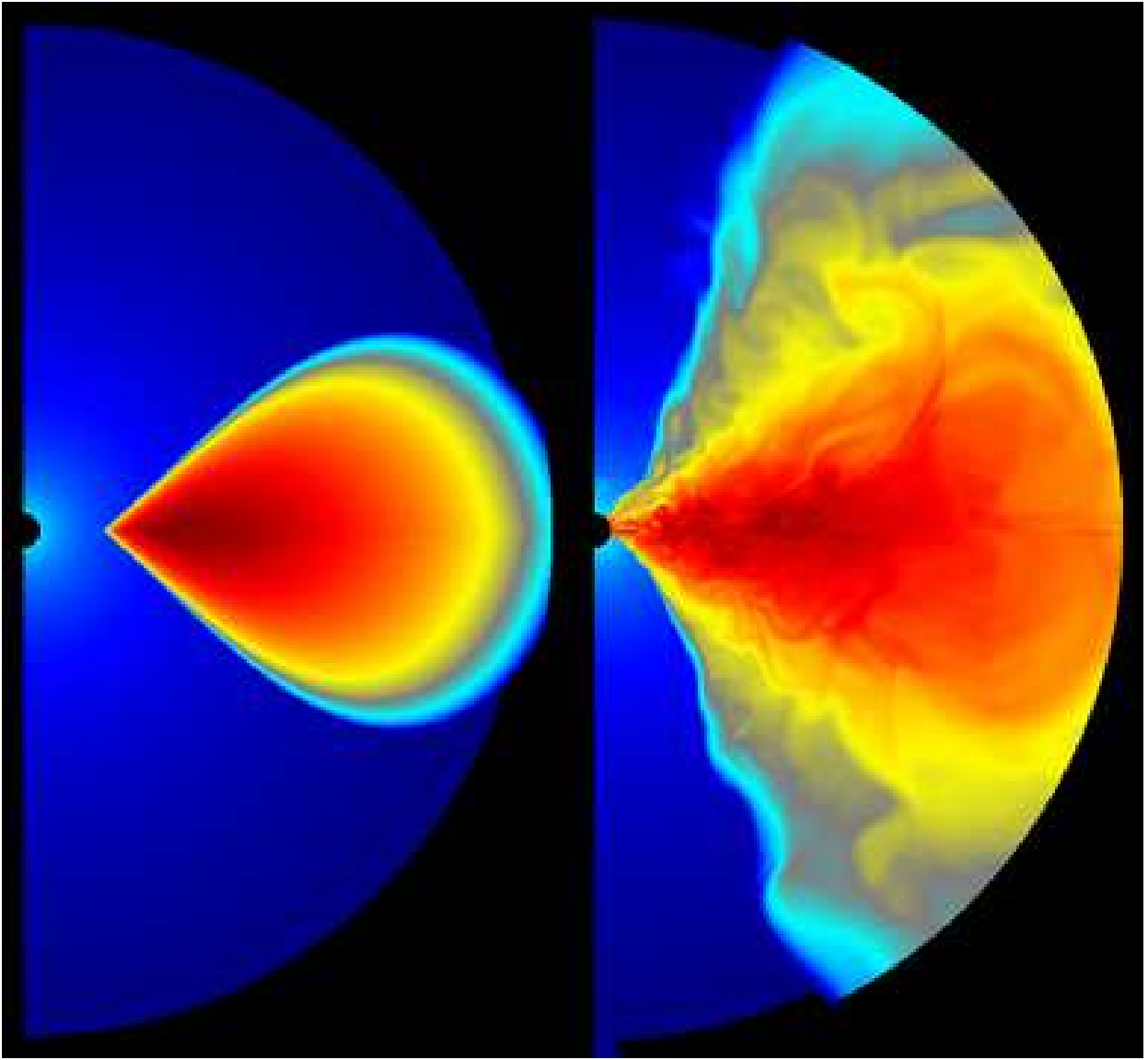}
\caption{Initial (left) and final (right) distribution of
$\log{\rho_0}$ in the fiducial model on the $r
\sin\theta-r\cos\theta$ plane.  At $t=0$ black corresponds to
$\rho_0\approx 4\times10^{-7}$ and dark red corresponds to
$\rho_0=1$.  For $t=2000$, black corresponds to $\rho_0\approx
4\times10^{-7}$ and dark red corresponds to $\rho_0=0.57$.  The
black half circle at the left edge is the black hole. }
\label{density}
\end{figure}

Figure~\ref{density} shows the initial and final density states
projected on the ($R,z = r\sin{\theta},r\cos{\theta}$)-plane. Color
represents $\log(\rho_0)$.  The initial density maximum is $1$ and
the minimum is $\approx 4 \times 10^{-7}$.  The final state contains
shocks driven by the interaction with the magnetic field, outflows
near the surface of the disk, and an evacuated ``funnel'' region
near the poles.

\begin{figure}
\centering
\subfigure{\includegraphics[width=3.2in,clip]{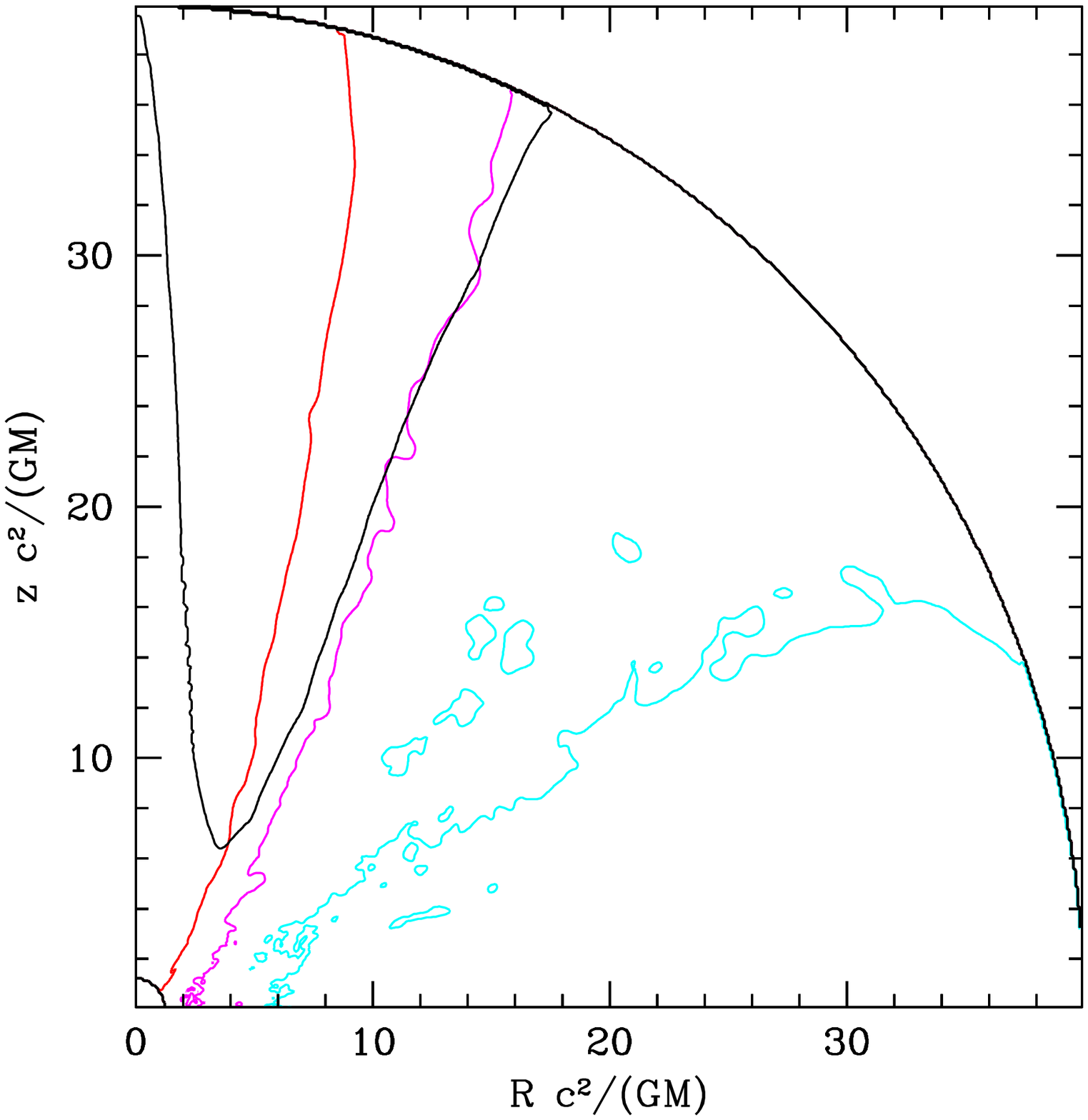}}
\subfigure{\includegraphics[width=3.2in,clip]{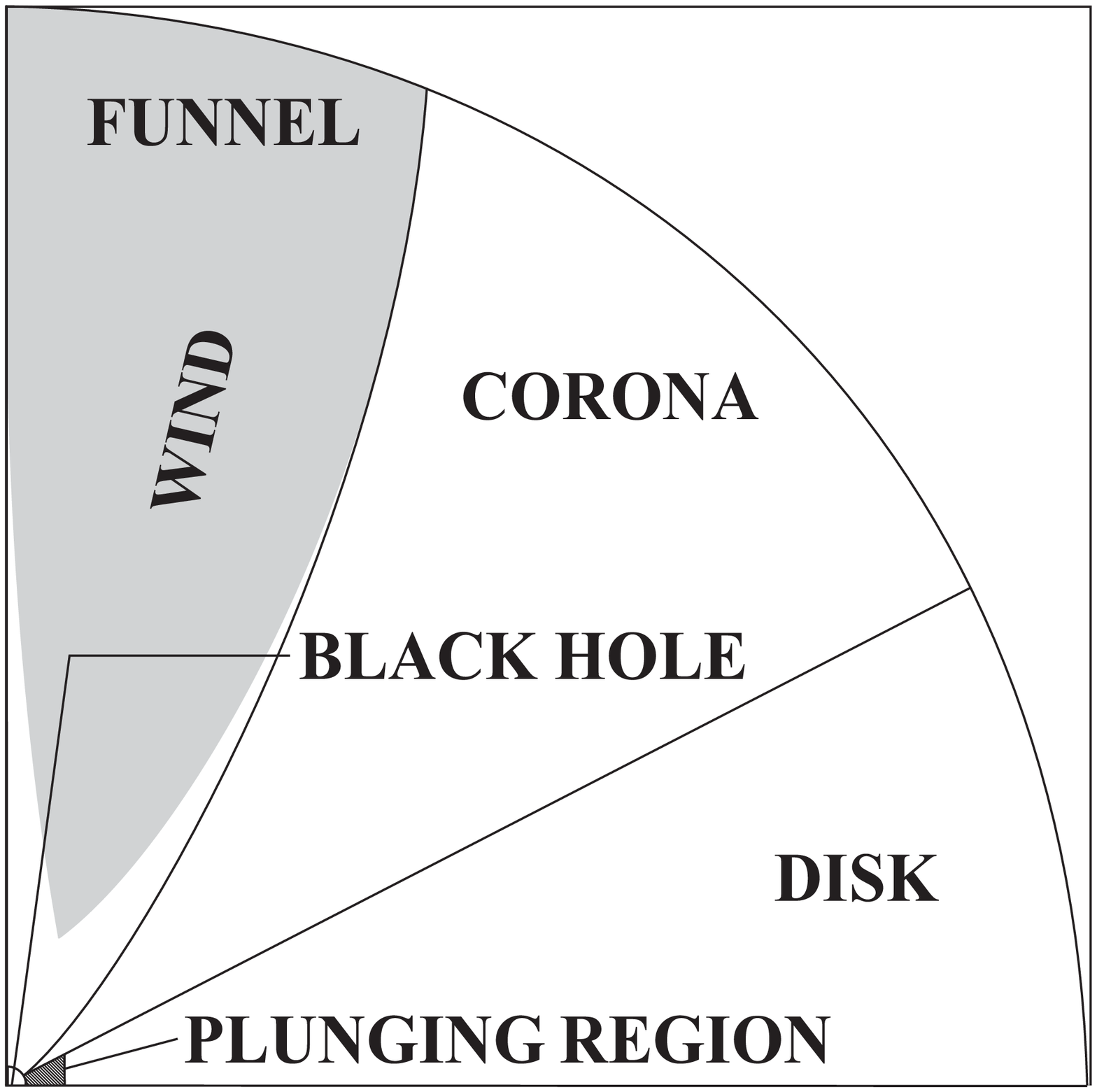}}

\caption{ (a) The distribution of $\beta$, $b^2/\rho_0$, and $u_t$
in the fiducial run, based on time and hemispherically averaged
data.  Starting from the axis and moving toward the equator: (1)
$u_t=-1$ contour shown as a solid black line; (2) $b^2/\rho_0=1$
contour shown as a red line; (3) $\beta=1$ contour shown as a
magenta line that nearly matches part of the $u_t=-1$ contour
line; and (4) $\beta=3$ contour is shown as cyan line. (b)
Motivated by the left panel, the right panel indicates the
location of the five main subregions of the black hole
magnetosphere.  They are (1) the disk: a matter dominated region
where $b^2/\rho_0 \ll 1$; (2) the funnel: a magnetically dominated
region around the poles where $b^2/\rho_0 \gg 1$ where the
magnetic field is collimated and twists around and up the axis
into an outflow; (3) the corona: a region in the relatively low
density upper layers of the disk with weak time-averaged poloidal
field; (4) the plunging region; and (5) the wind, which straddles
the corona-funnel boundary.  See section~\ref{fiducial} for a
discussion.} \label{cartoon}
\end{figure}

The left panel in figure~\ref{cartoon} indicates the relative
densities of internal, magnetic, and rest-mass energy.  The
magenta and cyan contours show the ratio of the average pressure
to average magnetic pressure, $\beta \equiv 2 \bar{p}/\bar{b^2}$.
The overbar indicates an average taken over $1000 < t < 2000$ and
over both hemispheres. The cyan contour indicates $\beta = 3$ and
encircles most of the high density, approximately Keplerian disk.
The magenta contour indicates $\beta = 1$.  The red contour
indicates where $\bar{b^2}/\bar{\rho_0}=1$.  Between the pole and
this contour the magnetic energy density exceeds the internal and
rest-mass energy density. The black contour surrounds a region,
extending to large radius, where $-u_t > 1$ and the flow is
directed outward (at large radius $-u_t$ asymptotes to the Lorentz
factor). That is, the particle energy-at-infinity is larger than
the rest-mass density: so the fluid is in a sense, unbound.  We
use the value of $u_t$ to estimate the radial component of the
3-velocity at infinity ($\tilde{v}^r$), which is independent of
the coordinate system.

The right panel in figure~\ref{cartoon} defines some useful
terminology inspired by the left panel, following \cite{dh03} and
\cite{hir03}. Moving from the axis to the equator, the ``funnel''
is the nearly evacuated, strongly magnetized region ($b^2\gg
\rho_0 + \IEDEN + p$), that develops over the poles.  The ``wind''
consists of a cone of material near the edge of the funnel that is
flowing outward with an asymptotic radial velocity of $\tilde{v}^r
\sim 0.75c$. Near the outer edge of our computational domain the
wind becomes marginally superfast. The ``corona'' lies between the
funnel and the disk and has $b^2/2\sim p$ except in strongly
magnetized filaments. In the ``disk'' $b^2/2 < p$ and the plasma
follows nearly Keplerian orbits. Finally, the ``plunging'' region,
which lies between the disk and the event horizon, contains
accreting material moving on magnetic field and pressure modified
geodesics.

\begin{figure}
\centering
\includegraphics[width=6.49in,clip]{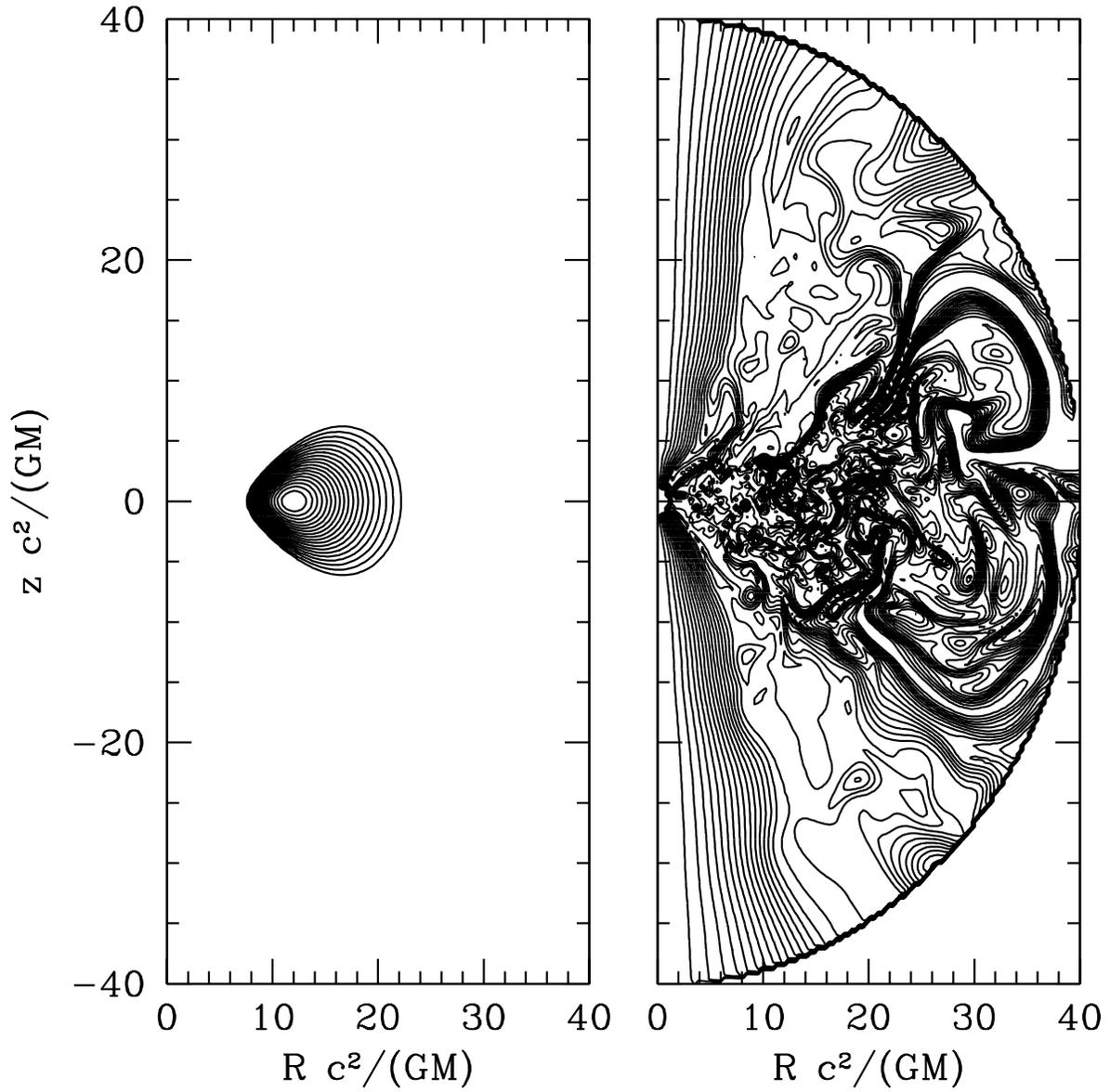}

\caption{ Initial (left) and final (right) distribution of
$A_\phi$.  Level surfaces coincide with magnetic field lines and
field line density corresponds to poloidal field strength.  In the
initial state field lines follow density contours if $\rho_0>0.2
\rho_{0,max}$. } \label{aphi}
\end{figure}

Figure~\ref{aphi} shows the evolution of the poloidal magnetic
field. The panels show contours of constant $A_\phi$, so the
density of contours is directly related to the poloidal field
strength, and the contours follow magnetic field lines.  The
contours are projected on the ($R = r\sin{\theta}$, $z =
r\cos{\theta}$)-plane, and show the initial and final state.  The
initial field is confined to a region much smaller than the torus
as a whole because field is introduced only in those portions of
the disk that have $\rho_0 > 0.2\rho_{max}$.  Notice that by the
end of the simulation the field has mixed in to the funnel region
and has a regular geometry there. In the disk and at the surface
of the disk the field is curved on the scale of the disk scale
height. The field strengths and geometries we see are consistent
with \citet{hir03}.  This includes the absence of disk to disk
field loops, and that the funnel field collimates instead of
connecting back into the disk (thus providing a means for the
outflow to escape to large radii).

\begin{figure}
\centering
\includegraphics[width=6.49in,clip]{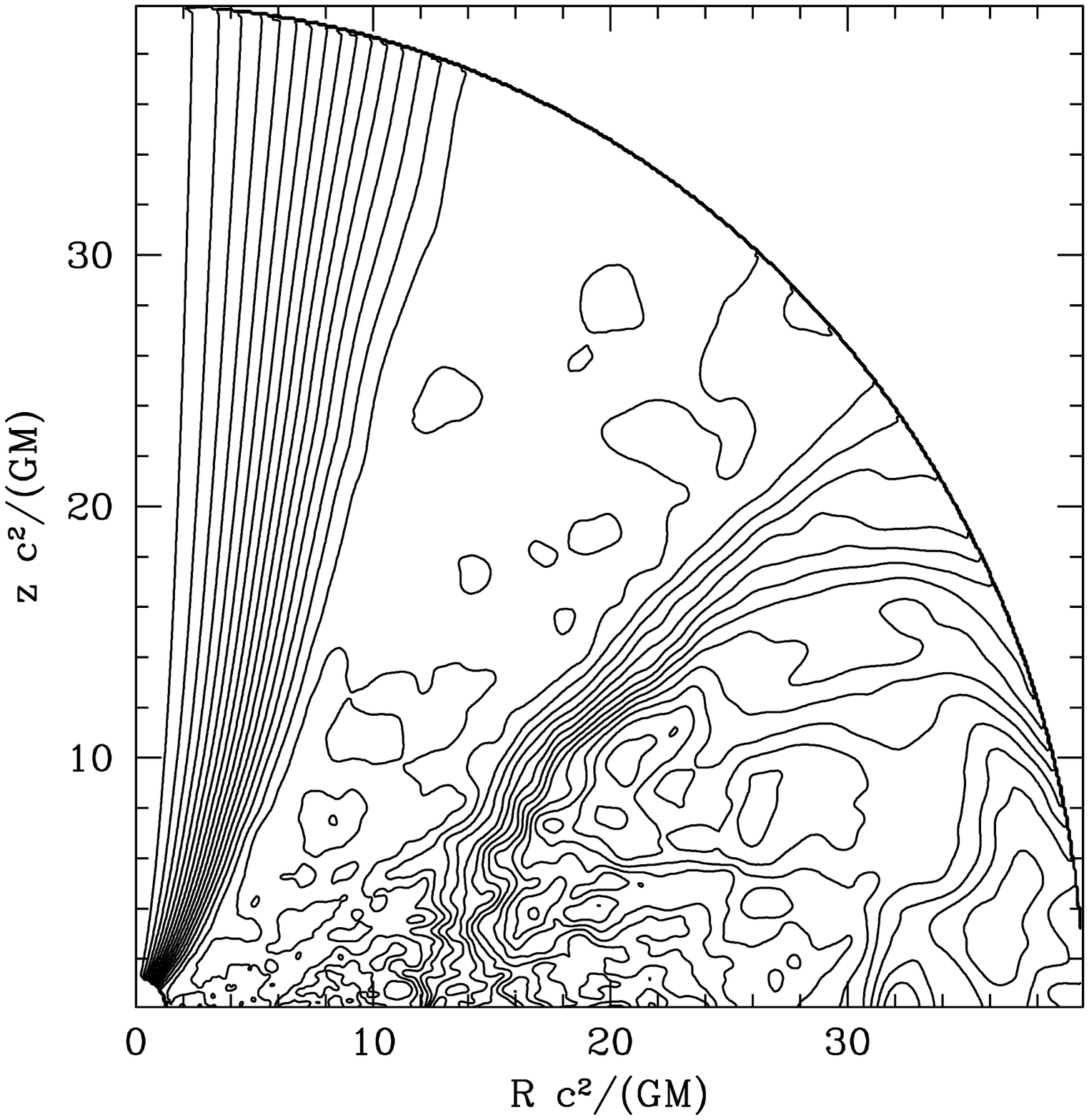}

\caption{ Contour plot of the time and hemispheric average of
$A_{\phi}$. Level surfaces coincide with magnetic field lines and
field line density corresponds to poloidal field strength. }
\label{tavgaphi}
\end{figure}

Figure~\ref{tavgaphi} shows contours of time and hemisphere
averaged $A_\phi$. The time averaged field is even more regular in
the funnel than the snapshot in Figure~\ref{aphi}. Time averaging
tends to sharply reduce the field strength in the corona and disk
because the field fluctuates in magnitude and direction there.

\begin{figure}
\centering
\includegraphics[width=6.49in,clip]{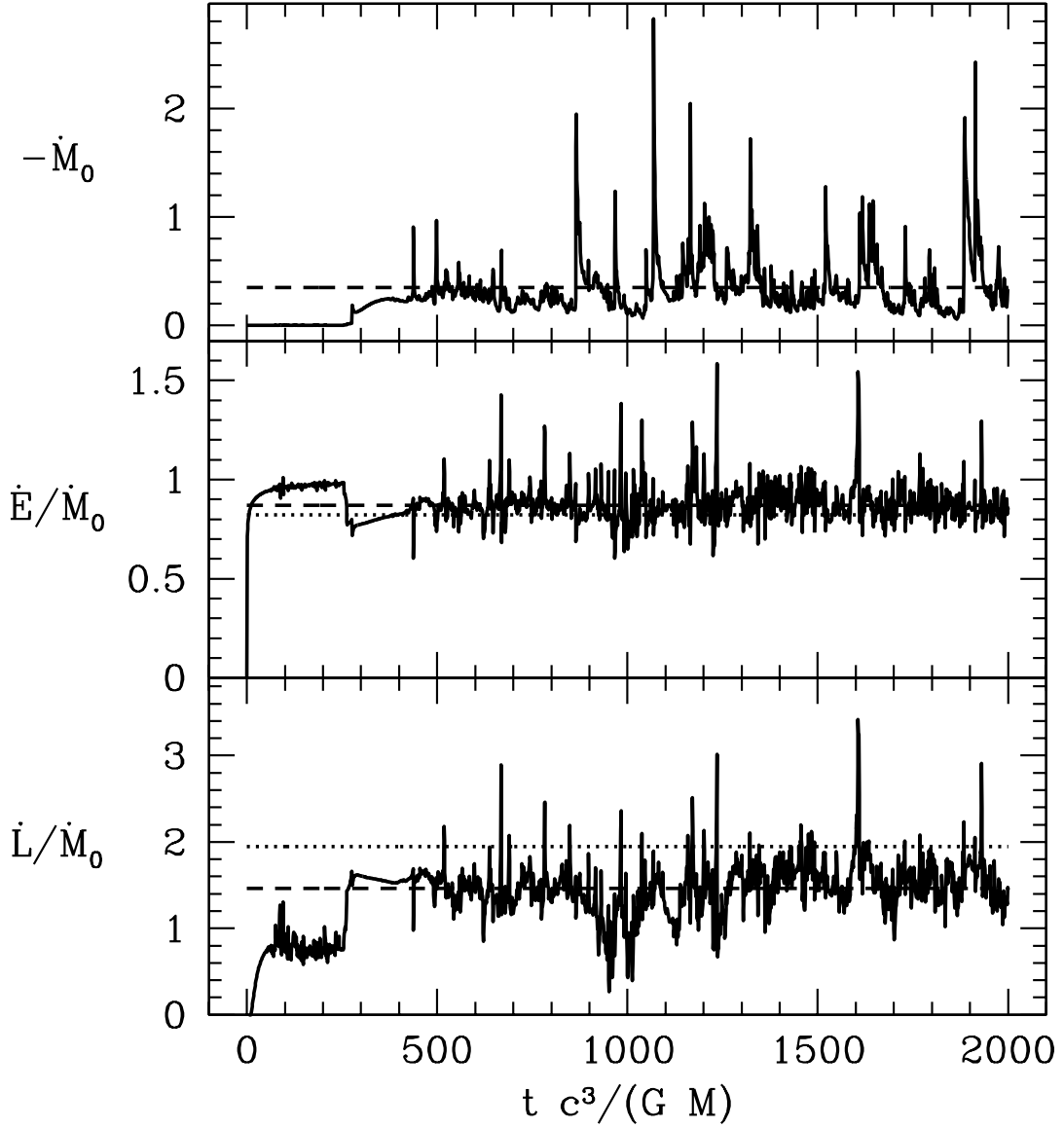}

\caption{ Evolution of rest-mass, energy, and angular momentum
accretion rate for our fiducial run of a weakly magnetized tori
around a black hole with spin $a=0.938$.  For $500 < t < 2000$ the
time average of these values is $\mflux\simeq 0.35$,
$\eflux/\mflux\simeq 0.87$, and $\lflux/\mflux \simeq 1.46$ as
shown by the dashed lines. The dotted lines show the classical
thin disk values ($\eflux/\mflux \simeq 0.82$ and $\lflux/\mflux
\simeq 1.95$). See section~\ref{fiducial} for a discussion.}
\label{3dotpanel}
\end{figure}

Figure~\ref{3dotpanel} shows the accretion rate of rest-mass
($\mflux$), energy per unit rest-mass ($\eflux/\mflux$), and angular
momentum per unit rest-mass ($\eflux/\mflux$) evaluated inside the
horizon at the inner boundary of the computational domain. For $500
< t < 2000$ the time average values are $\mflux\approx 0.35$,
$\eflux/\mflux\approx 0.87$, and $\lflux/\mflux\approx 1.46$. These
average values are shown as dashed lines.  The dotted lines show the
classical thin disk values $\eflux/\mflux \approx 0.82$ and
$\lflux/\mflux \approx 1.95$ obtained by setting these ratios equal
to respectively the specific energy and angular momentum of
particles on the ISCO.  The energy per baryon is therefore slightly
above the thin disk value, but the angular momentum per baryon is
significantly below the thin disk value.

It may be useful to recast the energy flux in terms of a nominal
``radiative efficiency''\footnote{Our evolution is nonradiative,
so the true radiative efficiency is zero.} $\eps = 1 -
\eflux/\mflux$. For the fiducial run $\eps = 13\%$, which is
slightly lower than the thin disk with $\eps = 18\%$.  This is
likely due to the high temperature of the flow. On the horizon
about $20$\% of the energy flux would vanish if we set the
internal energy to zero. The corresponding zero-temperature
efficiency ($1 + u_t$) would be 32\%.

\begin{figure}
\centering
\includegraphics[width=6.49in,clip]{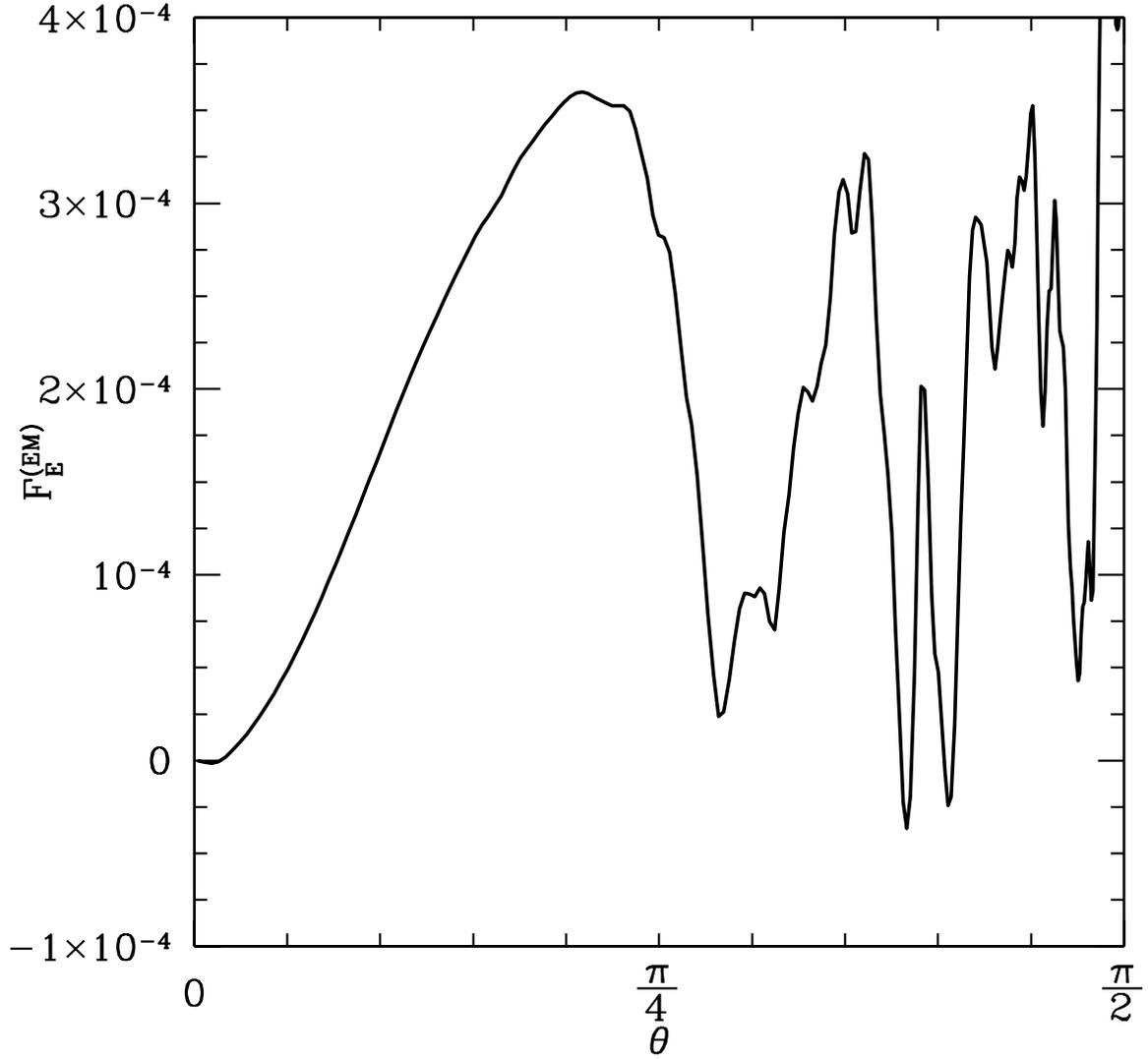}

\caption{ Electromagnetic energy flux density $\defluxem(\theta)$
on the horizon for the fiducial run, based on time and hemisphere
averaged data.  The mean electromagnetic energy flux is directed
outward.  See section~\ref{fiducial} for a discussion.}
\label{ervstheta}
\end{figure}

The chief object of our study is to measure the electromagnetic
luminosity of the hole.  The time and hemisphere averaged
electromagnetic energy flux on the horizon is shown in
figure~\ref{ervstheta}. In the funnel region the energy flux density is
outward, as predicted by the force-free model of BZ. We compute other
interesting quantities by integrating over the horizon and taking a time
average (for technical reasons we are using a less resolved time
sampling here than used to make figure~\ref{3dotpanel}, but the time
averages have fractional differences of only 10\%). We find
$\efluxem/\efluxma=-2.3\%$, where the energies per baryon are
$\efluxem/\mflux = -0.018$ and $\efluxma/\mflux = 0.77$.  It is useful
to define the ratio of electromagnetic luminosity to nominal accretion
luminosity $\tilde{L}=\efluxem/(-\eps\mflux)$.  We find
$\tilde{L}=16\%$.  Thus while the electromagnetic energy flux is
outward, it is a small fraction of the inward material energy flux and
the BZ luminosity is small compared to the nominal accretion luminosity.

A control calculation at $a = 0$ and a resolution of $256^2$ gives
$\efluxem/\efluxma=0.33\%$ and $\tilde{L}=-6.5\%$, where the
energies per baryon are $\efluxem/\mflux = 0.0032$ and
$\efluxma/\mflux = 0.95$. $\efluxem/\mflux > 0$ and $\tilde{L}<0$
are as expected, since the outward energy flux must vanish for a
nonrotating hole (i.e. the BZ effect is not operating). For our
sequence of models the BZ effect does not operate for $a\lesssim
0.5$ (see section~\ref{bhspin}).  The matter energy flux ratio may
be compared to the thin disk value of $\efluxma/\mflux=0.94$.

\subsection{Comparison with BZ}\label{bzcomp}

The BZ solution was reviewed in section~\ref{bzanalytic}.  BZ were
able to find steady force-free field solutions in the limit that
$a \ll 1$. Since the fiducial run has $a = 0.938$, we ran a
special $a = 0.5$ model for comparison with BZ.

The BZ solution was found in the force-free limit, so the first
question one might ask is whether any region of the model is
force-free.  To measure this we recall that in the force-free
limit
\begin{equation}
{T^{\mu\nu}}_{;\nu} = F^{\mu\nu} J_\nu = 0.
\end{equation}
So when the field is force-free the parameter
\begin{equation}
\zeta = \left| {F^{\mu\nu} J_\nu F_{\mu\kappa} J^\kappa\over
{J_\mu J^\mu F_{\kappa\lambda} F^{\kappa\lambda}}} \right|
\end{equation}
is small compared to 1.

\begin{figure}
\centering
\includegraphics[width=6.49in,clip]{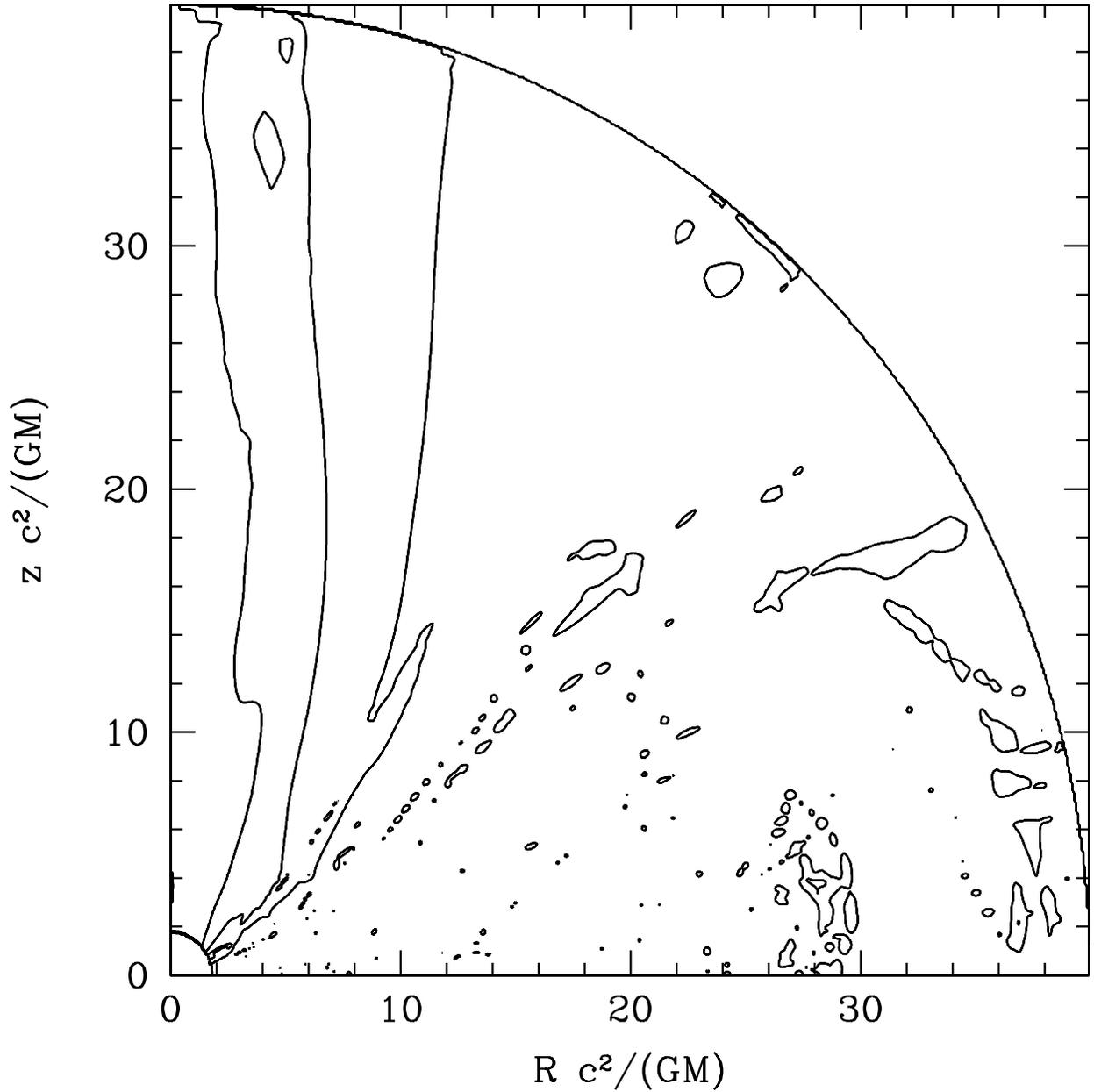}

\caption{ The run of the force-free parameter $\zeta$ for the $a =
0.5$ run; when $\zeta \ll 1$ the field is approximately
force-free.  The parameter has been time and hemisphere averaged.
The contours show (beginning from the pole and moving toward the
equator) $\zeta = 10^{-3}, 10^{-2}, 10^{-1}$.  The small closed
contours at large radius and close to the axis have
$\zeta=10^{-2}$. The small closed contours from the equator to
$\theta\sim\pi/4$ have $\zeta=10^{-1}$. See section~\ref{bzcomp}
for a discussion.}\label{zetaplot}
\end{figure}

Figure~\ref{zetaplot} shows the time and hemispherical averaged
$\zeta(r,\theta)$ from $t=1000$ to $t=2000$ for the $a=0.5$ model.
The contours show (beginning from the pole and moving toward the
equator) $\zeta = 10^{-3}, 10^{-2}, 10^{-1}$.  The entire funnel
region has $\zeta < 10^{-2}$ and is therefore effectively
force-free.  This is true in both a time-averaged and instantaneous
sense in the funnel for all our runs.  This opens the possibility
that the BZ solution describes the funnel.

\begin{figure}
\centering
\subfigure{\includegraphics[width=3.2in,clip]{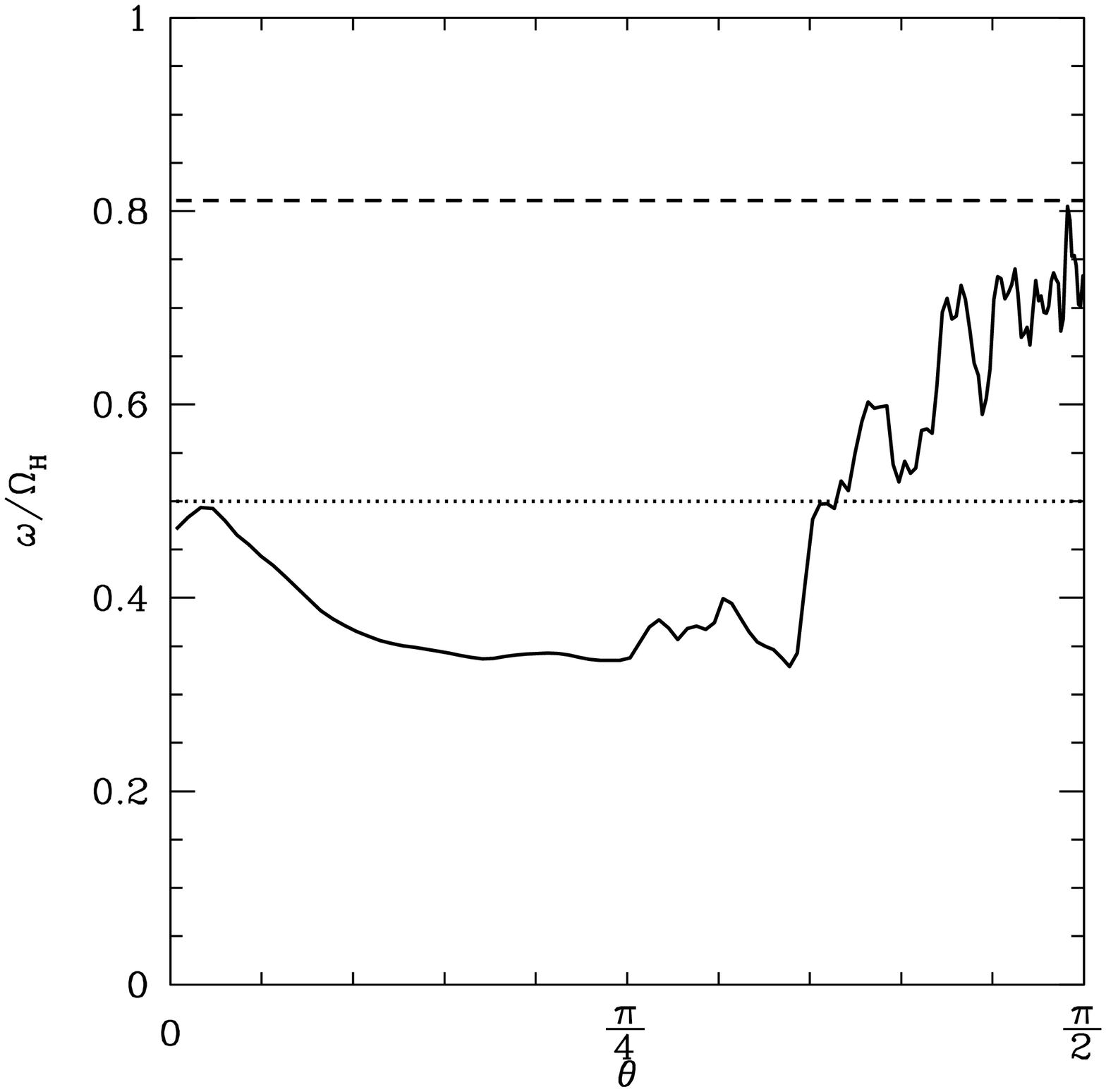}}
\subfigure{\includegraphics[width=3.2in,clip]{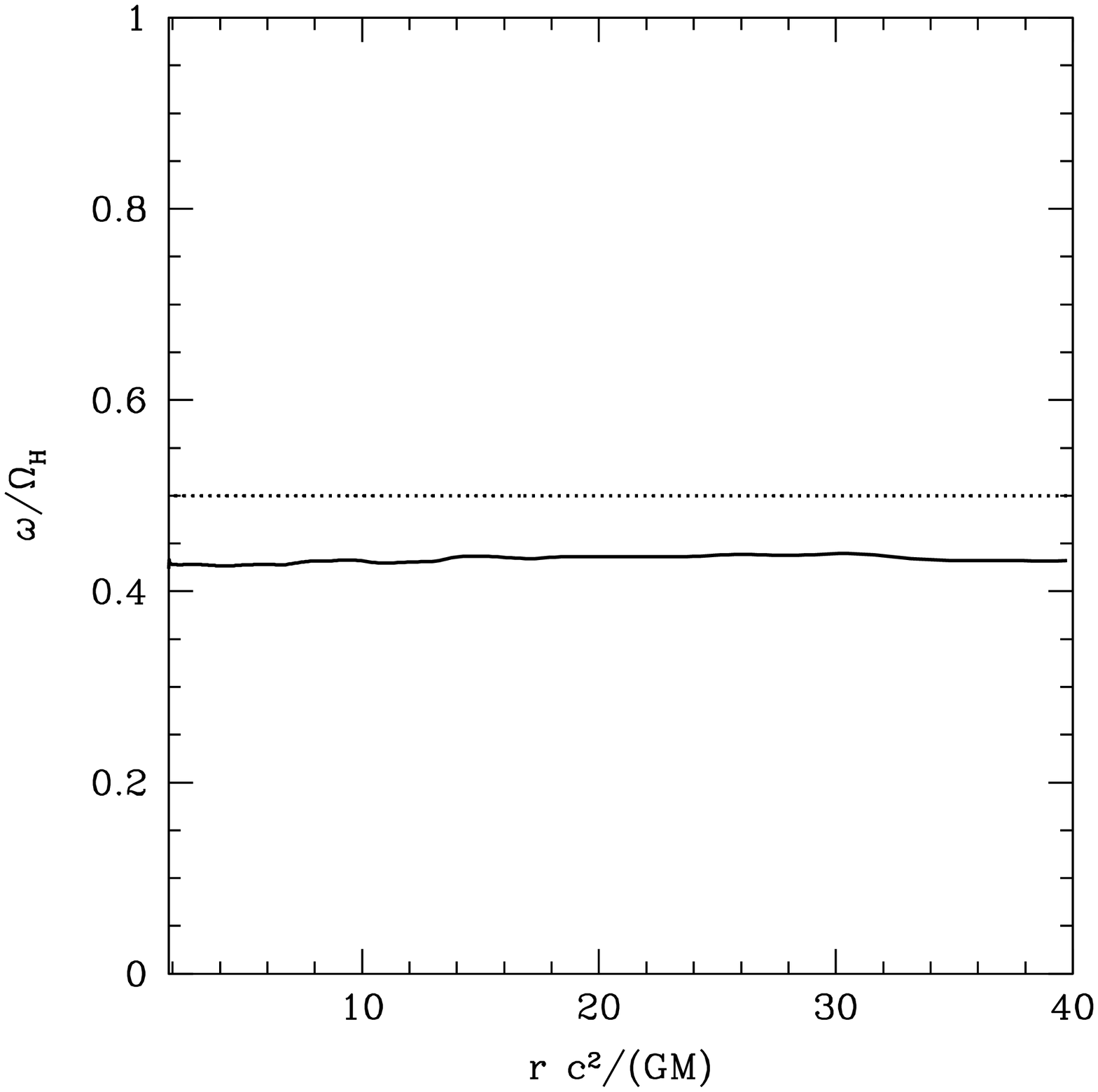}}

\caption{ Left panel: Magnetic field angular frequency on the
horizon relative to black hole rotation $\omega(\theta)/\Omega_H$.
The solid line indicates time and hemisphere averaged data from
our $a=0.5$ MHD integration.  The middle dotted line is the
prediction of the BZ model ($\omega/\Omega_H=1/2$).  The dashed
line (top) is the value predicted by the inflow model.  Right
panel: the run of field rotation frequency $\omega$ with radius
along a single field line that intersects the horizon at $\theta =
0.2$.  $\omega$ is constant to within 3\%, as expected for a
steady flow. See sections~\ref{bzcomp} and~\ref{inflowcomp} for a
discussion.} \label{omegavstheta}
\end{figure}
\epsscale{1.0}

\begin{figure}
\centering
\subfigure{\includegraphics[width=3.2in,clip]{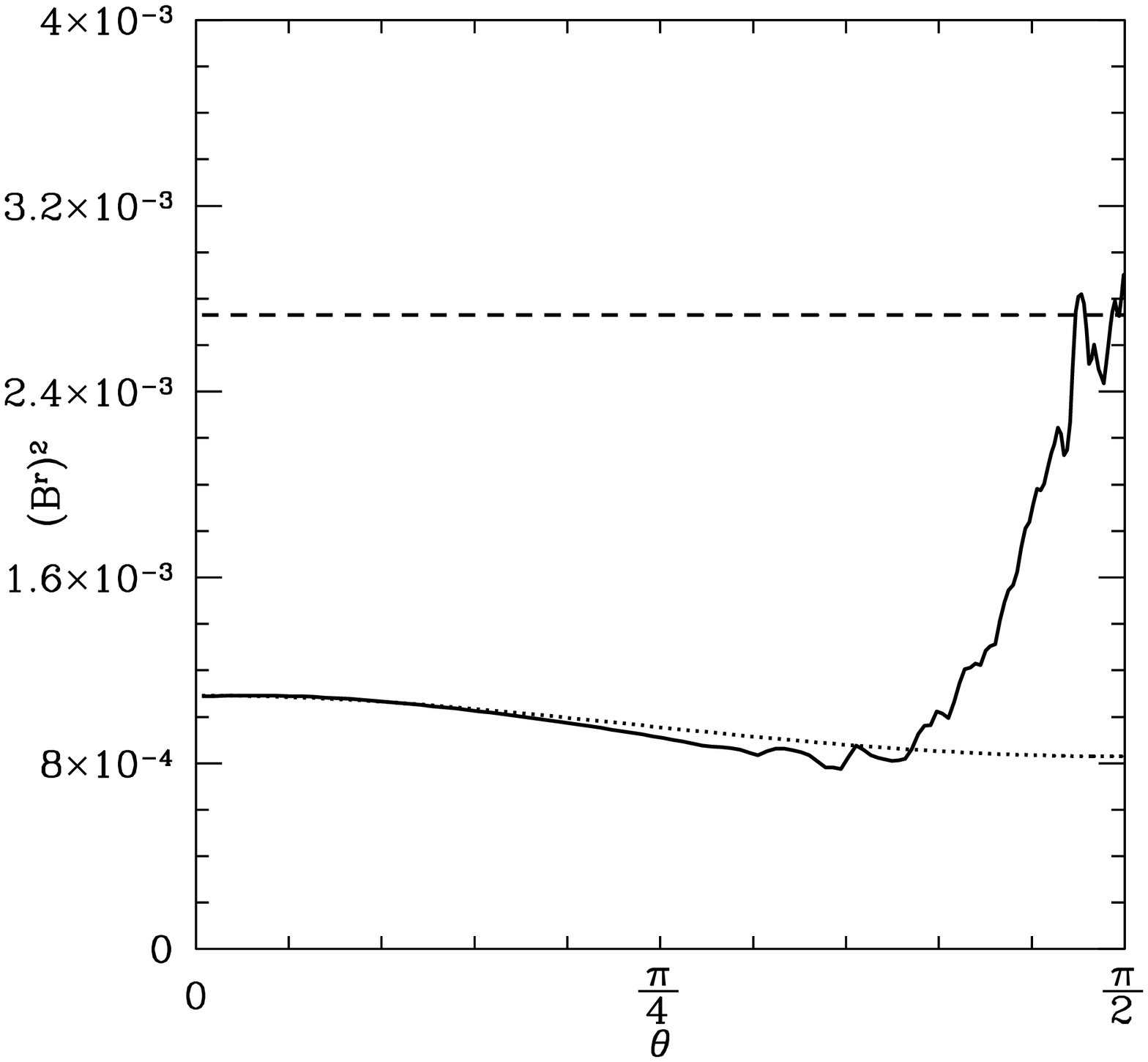}}
\subfigure{\includegraphics[width=3.2in,clip]{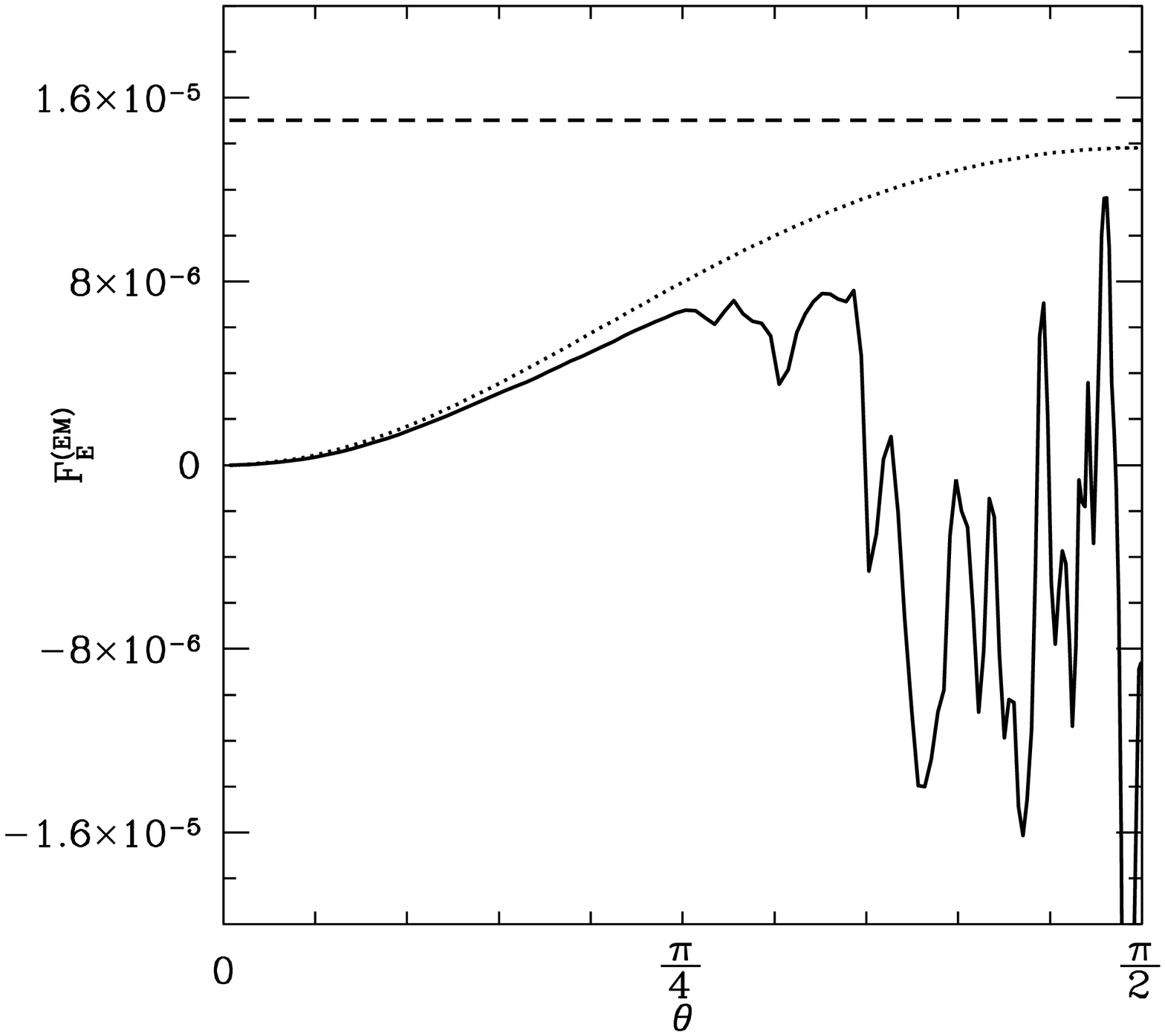}}

\caption{ (a) Square of radial field ($({B^r}(\theta))^2$) on the
horizon in the $a = 0.5$ MHD integration, from time and hemisphere
averaged data. Solid line is the field for our numerical model.
The dotted line shows the \citet{bz77} perturbed monopole solution
with the field strength normalized to the numerical solution at
the pole.  The dashed line is the inflow solution. (b)
Electromagnetic energy flux $\defluxem(\theta)$ on the horizon in
the $a = 0.5$ MHD integration, from time and hemisphere averaged
data. The solid line shows the numerical model, the dotted line
shows BZ's spun-up monopole solution, and the dashed line shows
the inflow solution.
 See sections~\ref{bzcomp} and~\ref{inflowcomp} for a discussion.}\label{bsqem}
\end{figure}

A key feature of the BZ model is that the field rotation frequency
$\omega \approx \Omega_H/2$ for $a \ll 1$ if the field has a
monopole geometry.  Figure~\ref{omegavstheta}a shows the ratio
$\omega/\Omega_H$ on the horizon. Within the force-free region,
which runs from $0 < \theta < 0.4$ on the horizon, the average
$\omega/\Omega_H \approx 0.45$. The small difference from the BZ
could be due to higher order terms in the expansion in $a$, but
\cite{kom01} has integrated the equations of force-free
electrodynamics for a monopolar field geometry and at $a = 0.5$
finds that $\omega$ rises from  $\approx 0.495 \Omega_H$ at the pole
to $\approx 0.51 \Omega_H$ at the equator, so this seems unlikely.
The difference is more likely due to small deviations from
force-free behavior (mass loading of field lines by the numerical
``floor'' on the density).

In an axisymmetric steady state both the force-free equations and
the MHD equations predict that the rotation frequency $\omega$ (and
other quantities) are constant along field lines.
Figure~\ref{omegavstheta}b shows the variation of $\omega$ with
radius along a field line that intersects the horizon at $\theta =
0.33$.  As expected $\omega \approx const.$, with a variation of
less than 3\% from maximum to minimum.

BZ's spun-up monopole model makes definite predictions about the
variation of $B^r$ and $\deflux$ on the horizon.
Figure~\ref{bsqem}a shows the variation in time and hemisphere
averaged $(B^r)^2$ and compares to BZ's monopole field
calculation. The single adjustable parameter of the model
normalizes the field strength.  We have set this normalization by
requiring that $(B^r)^2$ match at the pole. Evidently the
variation matches the BZ prediction closely even well outside the
force-free region at $\theta \approx 1.1$. Figure~\ref{bsqem}b
shows the variation in radial energy flux on the horizon as
predicted by the BZ model using the pole-normalized field.  Here
the match is quite close out to $\theta \approx \pi/4$. It is
slightly surprising that the BZ solution does so well even in
regions that are not force-free.  This is likely a result of
trans-field force balance and geometry controlling the
distribution of field on the horizon and hence the radial energy
flux.

To summarize: in our low spin numerical experiment the funnel is
approximately force-free within the funnel.  It is approximately
in a steady state and hence $\omega$ is approximately constant
along field lines.  Furthermore, $\omega$, $B^r$, and the radial
electromagnetic energy flux are all in good agreement with the
spun-up monopole force-free model on the horizon.  We have not
compared the entire funnel region with the monopole model because
the field is collimated there and not well described by the
monopole solution.

\subsection{Comparison to Inflow Solution}\label{inflowcomp}

The inflow solution of \cite{gam99} considers a near-equatorial
stationary MHD inflow in the plunging region, reviewed in
section~\ref{inflowanalytic}. Here we compare the inflow models
with the fiducial model. Unlike the funnel, the plunging region is
rapidly fluctuating, so we expect the inflow model to match only
the time-averaged data from the simulation.

The inflow model has two free parameters: the field strength and the
accretion rate.  The field strength we match by finding the parameter
that gives the best fit to the mean magnetic energy density between the
ISCO and the event horizon.  The rest-mass flux is chosen to agree with
the time-averaged data from the simulation.  The ratio of the field
strength to the square root of the accretion rate is a dimensionless
parameter that controls the solution; in the units of \cite{gam99},
where $2 \pi \rho_0 u^r \detg = -1$, we use $F_{\theta\phi} = 1.09$ for
the comparison model.

\epsscale{0.7}
\begin{figure}
\centering
\includegraphics[width=6.49in,clip]{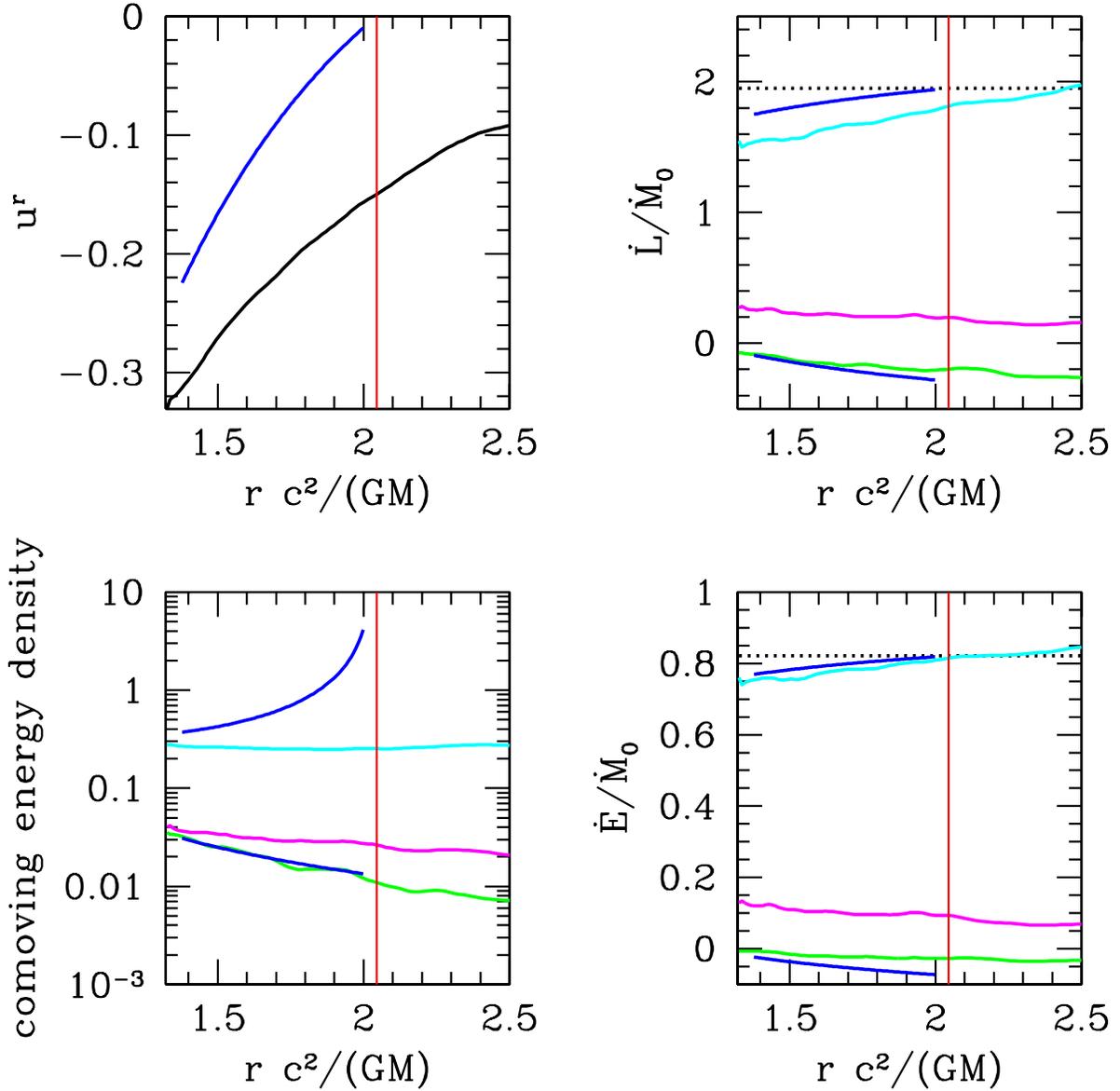}

\caption{ A comparison of the time-averaged fiducial model near
the equator (within $\theta = \pi/2 \pm 0.3$) with the inflow
solution of \cite{gam99}.  In the right two panels the black
dotted line is the thin disk value.  In all cases the red vertical
line is the location of the ISCO. The black line for the upper
left panel is the numerical result. For the other three panels,
the particle term is shown in cyan, the internal energy term is
shown in magenta, and the electromagnetic term is shown in green.
The blue line in each plot represents the inflow model result.
Notice that the run of density with radius shows no feature at the
ISCO.  See the Section~\ref{inflowcomp} for discussion. }
\label{inflowcompplot}
\end{figure}

Figure~\ref{inflowcompplot} shows a comparison of $u^r$,
$\lflux/\mflux$, comoving energy densities ($\rho_0$, $b^2/2$, and
$\IEDEN$), and energy fluxes ($\eflux/\mflux$) in the inflow
solution.  The comparison data from the fiducial run has been
averaged over $|\theta - \pi/2| < 0.3$ and $500 < t < 2000$.  Each
panel in the figure contains a vertical line at the ISCO.

The upper left panel compares the radial component of the
four-velocity (in KS and BL coordinates) in the inflow and numerical
solutions.  The substantial differences are due to the finite
temperature of the flow; the inflow solution is cold by assumption.
Radial pressure gradients in the numerical model (which are absent
in the inflow solution) begin to accelerate material inward outside
the ISCO, and the flow becomes supersonic near the ISCO.

The upper right and lower right panels show components of the
energy and angular momentum flux from the simulation and inflow
solutions.  The dashed horizontal line in each case shows the
values expected for a thin disk; the inflow solution is
constrained to match the thin disk at the ISCO.  The cyan lines
show $u_\phi$ (upper panel) and $u_t$ (lower panel) from the
simulation, while the blue lines show the prediction from the
inflow model.  The energy flux matches rather well (although
notice that this is only a small fraction of the energy flux),
while the angular momentum is overestimated; in the simulation the
plasma has sub-Keplerian angular momentum by the time it reaches
the ISCO.

The electromagnetic components of the normalized angular momentum
flux ($b^2 u_\phi/\rho_0 - b^r b_\phi/(\rho_0 u^r)$) and energy flux
($b^2 u_t/\rho_0 - b^r b_t/(\rho_0 u^r)$) are also shown in the
upper and lower right panels of figure~\ref{inflowcompplot} (green
line $\equiv$ simulation, blue line $\equiv$ inflow solution). The
inflow solution matches well, although it tends to overestimate the
magnitude of the outward directed energy flux.

The magenta lines in the upper and lower right panels show the
internal energy component of the normalized angular momentum flux
($(\IEDEN + p) u_\phi/\rho_0$) and energy flux ($-(\IEDEN + p)
u_t/\rho_0$). This component of the fluxes is zero by assumption in
the inflow solution, and it is evidently an important component of
the fluxes in our thick disk simulations.  This leads to large
corrections to the angular momentum and energy fluxes; the total
normalized angular momentum flux is significantly smaller than the
thin disk prediction, while the energy flux is, seemingly by
conspiracy, very close to the thin disk.

The lower left panel shows the rest-mass density from the inflow
solution (upper blue line) and from the simulation (cyan line).
The mass flux in the inflow solution is normalized so that it
matches the simulation mass flux. Since mass flux is approximately
constant with radius, the run of density is directly related to
the run of $u^r$. What is remarkable here is that there is no
feature in the simulation $\rho_0$ near the ISCO.  In fact it is
nearly constant from well outside the ISCO in to the event
horizon.  The surface density varies smoothly as well.  This
confirms the point made by \cite{kh02} in their pseudo-Newtonian
solution: {\it there is no sharp feature at the ISCO.} This has
implications for iron line profiles, as discussed by \cite{rb97}.

The lower left panel also shows the run of internal energy density in
the simulation (it is zero by assumption in the inflow solution).
Again, there is no sharp feature at the ISCO, just a gentle rise inward
toward the event horizon.  Because the density is nearly constant with
radius this implies that entropy is increasing inward.  Therefore there
is some dissipation of kinetic or magnetic energy into internal energy
in the inflow region.

The lower left panel of figure~\ref{inflowcompplot} shows the run of
magnetic energy density $b^2/2$ in the inflow solution (lower blue
line) and simulation (green line).  The normalization of the inflow
magnetic energy is a parameter, but its radial slope is not.

Finally, the inflow solution predicts that $\omega = \Omega_{ISCO}$.
Figure~\ref{omegavstheta}a shows the run of $\omega/\Omega_H$ on the
horizon for the $a = 0.5$ model.  The dashed line shows the ISCO
value of $\omega/\Omega_H$. At the equator the time-averaged
numerical value lies within about $10\%$ of the ISCO value: the
numerical average $\omega/\Omega_H = 0.685$, while the
$\Omega_{ISCO}/\Omega_H = 0.8136$ at the ISCO. In the $a = 0.938$
run the numerical average $\omega/\Omega_H =0.681$, while
$\Omega_{ISCO}/\Omega_H = 0.745$ at the ISCO.

To sum up, the inflow model does a surprisingly good job of
matching some aspects of the time-averaged simulation.  It does
not match the profile or boundary condition at the ISCO for the
radial velocity or the total angular momentum and energy fluxes,
because the simulation flow is hot, while the inflow solution has
zero temperature by assumption.

What is most surprising is that the energy per baryon accreted in
the numerical model matches the thin disk prediction.  The inflow
model predicts that the energy per baryon accreted should be {\it
lower} than the thin disk prediction, enhancing the nominal
accretion efficiency \citep{gam99,krolik99,ak00}.  The difference
is apparently due to the finite temperature of the numerical model
and the consequent change in boundary conditions at the ISCO.
These boundary conditions evidently adjust themselves to maintain
the energy flux at the thin disk value.  The angular momentum flux
is affected by the field, however, with the specific angular
momentum of the accreted material in the fiducial run about $25\%$
lower than the thin disk.

\section{Parameter Study}\label{parameterstudy}

Our numerical model has a number of physical and numerical parameters.
Here we check the sensitivity of the model to: (1) black hole spin
parameter $a$; (2) initial magnetic field geometry and initial magnetic
field strength; and (3) numerical parameters such as (a) location of the
inner boundary ($R_{in}$); (b) outer radial ($R_{out}$) boundary; (c)
radial and $\theta$ resolution, including the coordinate parameter $h$;
and (d) parameters describing the density and internal energy floors.

\subsection{Black Hole Spin}\label{bhspin}

The fiducial run has a rather low outgoing electromagnetic energy
flux compared to the ingoing matter energy flux. It is possible
that this varies sharply with black hole spin and that more
rapidly rotating holes exhibit much larger electromagnetic
luminosity.  We have performed a survey over $a$, keeping all
parameters identical to those in the fiducial run, except that the
resolution is lowered to $256^2$ and the location of the pressure
maximum is adjusted to keep $H/R \approx const$.

\begin{figure}
\centering
\includegraphics[width=6.49in,clip]{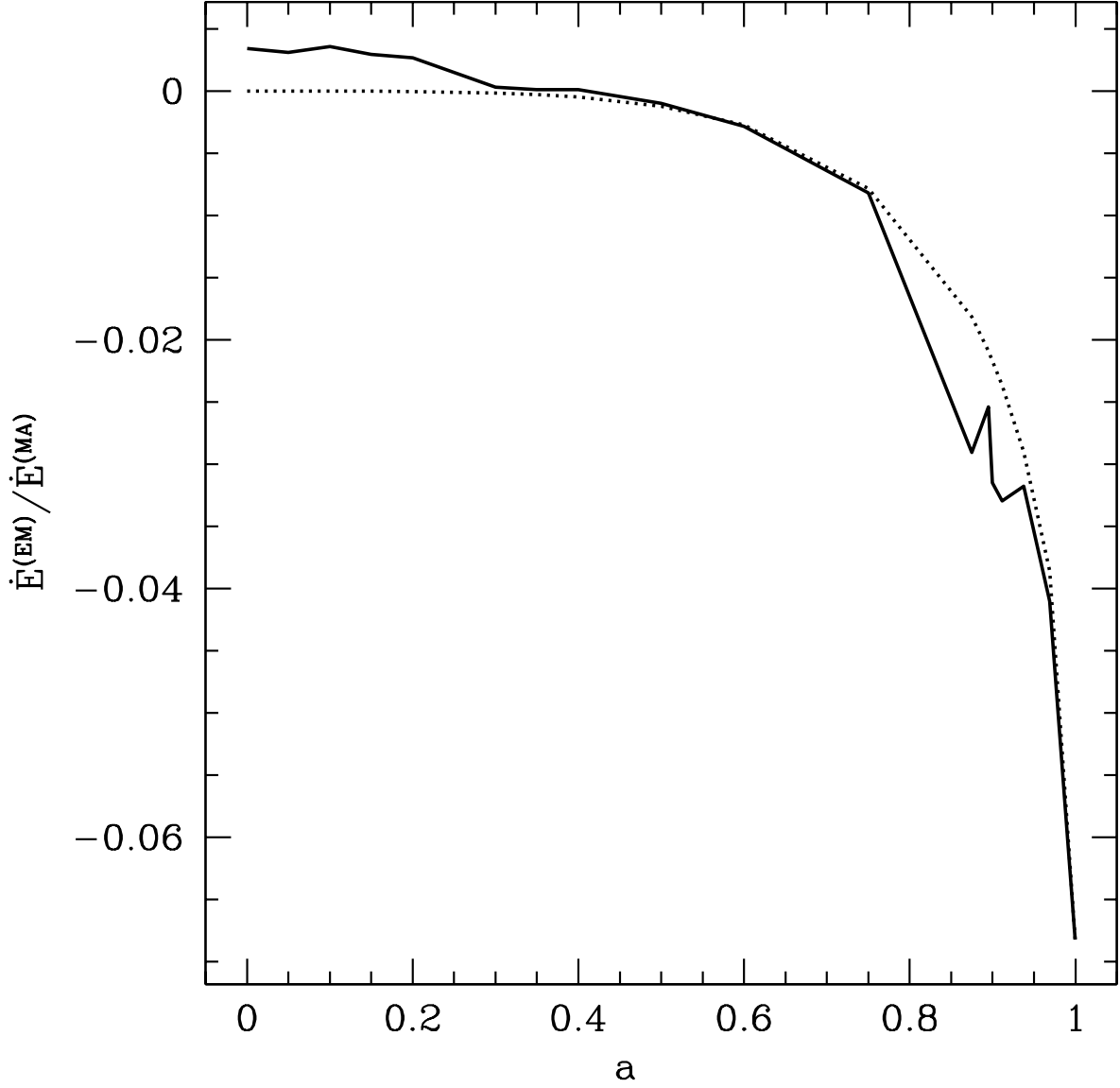}

\caption{ The ratio of electromagnetic to matter energy flux on
the horizon.  The solid line indicates numerical data while the
dotted line indicates a best fit of $\efluxem/\efluxma = -0.068 (2
- r_+)^2$.  See section~\ref{bhspin} for a discussion.}
\label{emtomatfull}
\end{figure}

\begin{deluxetable}{lllllll}
\tablewidth{0pt}

\tablecaption{ Black Hole Spin Study \label{tbl2}}

\tablehead{

\colhead{$a$}

& \colhead{$10^4\times\efluxem/\efluxma$}

& \colhead{$\eflux/\mflux$}

& \colhead{$\lflux/\mflux$}

& \colhead{$\dot{a}/\mflux$}

& \colhead{$\mflux$}

& \colhead{$\tilde{L}$}

}

\startdata

-0.938  &   105 &   0.958   &   3.806   &   -5.583  &   -0.908 & -0.24  \\

0.000   &   34.4  &   0.950   &   3.068   &   -3.049  &   -0.870  & -0.065\\

0.050   &   31.2  &   0.952   &   3.025   &   -2.921  &   -0.709  & -0.062\\

0.100   &   35.8  &   0.948   &   2.896   &   -2.713  &   -0.767  & -0.066\\

0.150   &   29.7  &   0.949   &   2.881   &   -2.597  &   -0.796  & -0.055\\

0.200   &   26.9  &   0.948   &   2.817   &   -2.439  &   -0.776  & -0.050\\

0.250   &   9.17   &   0.946   &   2.749   &   -2.302  &   -0.747 & -0.016 \\

0.300   &   3.30   &   0.937   &   2.759   &   -2.217  &   -0.571 & -0.0049 \\

0.350   &   1.32   &   0.933   &   2.605   &   -1.975  &   -0.620 & -0.0018 \\

0.400   &   1.15   &   0.937   &   2.763   &   -1.986  &   -0.241 & -0.0017 \\

0.500   &   -9.85  &   0.933   &   2.583   &   -1.665  &   -0.252 & 0.014 \\

0.600   &   -28.5 &   0.929   &   2.489   &   -1.347  &   -0.318  & 0.037\\

0.750   &   -81.8 &   0.908   &   2.150   &   -0.808  &   -0.276  & 0.083\\

0.875   &   -291    &   0.852   &   1.440   &   -0.152  &   -0.170 & 0.17 \\

0.895   &   -254    &   0.891   &   1.723   &   -0.204  &   -0.215 & 0.20 \\

0.900   &   -315    &   0.882   &   1.674   &   -0.118  &   -0.193 & 0.24 \\

0.938   &   -318    &   0.856   &   1.396   &   0.067   &   -0.203 & 0.23 \\

0.969   &   -410    &   0.869   &   1.374   &   0.217   &   -0.172 & 0.27 \\

\enddata
\tablecomments{
All models same as fiducial except at a resolution of $256^2$ and
$r_{max}$ is used to keep $H/R\sim constant$.  These values can be
compared to Tables 3 and 4. The efficiency is $1-\eflux/\mflux$. A
positive $\dot{a}/\mflux$ corresponds to a spindown of the black hole
since $\mflux < 0$.
}

\end{deluxetable}

The results are shown in figure~\ref{emtomatfull} and described in
table~\ref{tbl2}.  The figure shows the measured ratio of
electromagnetic to rest-mass energy flux; the dashed line shows a
fit
\begin{equation}\label{fittotal}
\frac{\efluxem}{\efluxma} \approx -0.068 (2 - r_+)^2.
\end{equation}
This fit applies only to this particular sequence of models;
models with different initial field geometries give different
results, as we shall see below. For all $a > 0$ we find $\efluxem
> 0$ in the funnel. For $a < 0.5$ this outward funnel flux is
balanced by an inward electromagnetic energy flux near the
equator.  For our most extreme run with $a = 0.969$ the outward
electromagnetic flux is still dominated by the inward particle
flux.  The ratio of electromagnetic luminosity to nominal
accretion luminosity is $\tilde{L}=27\%$, so the nominal accretion
luminosity dominates over the BZ luminosity.

The accretion rate of angular momentum is also a strong function of
spin.  As discussed in \cite{gsm2004}, accretion flows around rapidly
spinning holes have $d a/d t < 0$.  Our fiducial model, in fact, is
spinning down.  Previous estimates suggested that spin equilibrium is
reached at $a \sim 0.998$ \citep{thorne74}.  Our models reach spin
equilibrium at $a \sim 0.92$.

The variation of field strength and geometry with black hole spin
is also of interest.  To measure variation of field strength, we
probe the flow near four locations: 1) in the funnel near the
horizon (``funnel/horizon''); 2) in the plunging region near the
horizon (``plunging/horizon''); 3) at the ISCO; and 4) at the
pressure maximum. We then take a time and spatial average of the
comoving electromagnetic energy density $b^2/2$ over a small
region near each of these locations.  The ratio of
$b^2$(funnel/horizon) to $b^2$(plunging/horizon) changes from
$0.43$ at $a = 0$ to $0.74$ at $a = 0.938$.  The ratio of
$b^2$(funnel/horizon) to $b^2$(ISCO) varies from $2.53$ at $a = 0$
to $2.14$ at $a = 0.938$.  The ratio $b^2$(funnel/horizon) to
pressure maximum varies from $4.8$ at $a = 0$ to $15.7$ at $a =
0.938$.  In summary, the field strength increases from the ISCO to
the horizon by a factor of $\sim 3$ at $a=0$ and by a factor of
$\sim 6$ at $a=0.938$, and on the horizon is slightly larger at
the equator than at the poles by a factor of $\sim 2$. Only the
ratio of $b^2$(pressure maximum) to other locations in the
plunging region or at the horizon depends strongly on black hole
spin.

Our observed increase in horizon field strength with black hole spin
agrees with results reported by \cite{dhk03}.  We see no sign of the
expulsion of flux from the horizon reported by \cite{bj85}, who find
that the flux through one hemisphere of the horizon, due to external
sources and calculated in axisymmetry using vacuum electrodynamics,
vanishes when the spin of the hole is maximal. It is possible that we
have not gone close enough to $a = 1$ to observe this effect.

To investigate the variation of field geometry in the funnel region with
$a$ we trace field lines from $\theta_{in}$ on the horizon to
$\theta_{out}$ on the outer boundary and define a {\it collimation
factor} $\theta_{in}/\theta_{out}$.  The collimation factor is similar
for all field lines in the funnel region.  It reaches a minimum of
$\approx 5/2$ for the fiducial run, and rises to nearly $2$ for $a = 0$
and again to nearly $2$ for $a \sim 1$.  The collimation factor depends
on the location of the outer boundary; for models with $R_{out} = 400$
the collimation factor is $10$ and the field lines are nearly
cylindrical at the outer boundary.

We have also studied the variation of the field rotation frequency
$\omega$ in the funnel.  $\omega/\Omega_H$ varies weakly with $a$,
from $ 0.53$ at $a = 0.25$ to $0.45$ at $a = 0.938$, consistent with
the hypothesis advanced by \cite{membranebook} that $\omega/\Omega_H
\approx 1/2$.

\subsection{Field Geometry and Strength}

The outcome of the simulation may also depend on the field
geometry and strength in the initial conditions.  This seems more
likely for axisymmetric models such as ours where the evolution
may retain a stronger memory of the initial conditions than
comparable three dimensional models.

\begin{deluxetable}{llllllll}
\tablewidth{0pt}

\tablecaption{ Field Strength and Geometry Study\label{tbl3}}

\tablehead{

\colhead{Field Geometry $A_{\phi}$}

& \colhead{$\beta$}

& \colhead{$\efluxem/\efluxma$}

& \colhead{$\eflux/\mflux$}

& \colhead{$\lflux/\mflux$}

& \colhead{$\dot{a}/\mflux$}

& \colhead{$\mflux$}

& \colhead{$\tilde{L}$}

}

\startdata
 $A^0_{\phi}$                  & $100$  & $-0.0312$ & $0.856$ & $1.40$  & $0.0674$ & $-0.203$ & $0.21$ \\
 $A^0_{\phi}$                  & $500$  & $-0.0115$ & $0.879$ & $1.94$  & $-0.293$ & $-0.0474$ & $0.085$ \\
 $A^0_{\phi}\sin{(\log{(r/h)})}$ & $100$  & $-0.0355$ & $0.892$  & $1.24$  & $0.278$ & $-0.541$ & $0.42$ \\
 $A^0_{\phi}|\sin(2\theta)|$   & $100$  & $-0.0112$ & $0.888$ & $1.91$  & $-0.299$ & $-0.0746$ & $0.083$ \\
 $r\sin{\theta}$               & $100$  & $-0.147$  & $0.773$ & $-0.997$& $1.807$  & $-1.769$ & $0.79$ \\
 $r\sin{\theta}$               & $400$  & $-0.157$  & $0.813$ & $0.0617$& $1.184$  & $-0.715$ & $0.67$ \\
\enddata
\tablecomments{
$A^0_{\phi}$ is the fiducial model field geometry and $\beta=100$ is the
fiducial ratio of gas to magnetic pressure. The $r\sin{\theta}$ field
geometry is a uniform vertical field model with $\beta$ set by disk
values at the equator.  All other model and numerical parameters are as
in the fiducial model except that the resolution is $256^2$.  The
efficiency is $1-\eflux/\mflux$.  A positive $\dot{a}/\mflux$
corresponds to a spindown of the black hole because $\mflux < 0$.
}

\end{deluxetable}

We begin by investigating the dependence of outcome on initial
field strength, parameterized by $\beta \equiv
p_{gas,max}/p_{mag,max}$ (notice that the two maxima never occur
at the same location in space, so this ratio varies over a wide
range when evaluated at individual locations in the disk).  We
consider models with $\beta=(100,500)$ and find a weak dependence
on $\beta$. For the $\beta=100$ model (the fiducial model at a
resolution of $256^2$) we find $\omega/\Omega_H\approx 0.45$,
$\efluxem/\efluxma\approx -3.1\%$, and $\tilde{L}=21\%$.
$\beta=500$ leads to $\omega/\Omega_H\approx 0.42$,
$\efluxem/\efluxma=-1.2\%$, and $\tilde{L}=8.5\%$. Notice that a
higher spatial resolution is required to fully resolve weak field
models, although all runs in this comparison were done at $256^2$;
the decrease in electromagnetic energy extracted at $\beta = 500$
may therefore be due to resolution.

We also vary the field geometry from the single loop used in our
fiducial model, which has vector potential $A_{\phi}\propto
MAX(P/P_{max} -0.2, 0)$.  We do this by multiplying the vector
potential by $\sin(\log(r/h))$ or $|\sin(2\theta)|$.  The former
decompresses the field lines at the inner radial edge giving a
field strength that is more uniform around the loop (for an
extended disk this would yield a sequence of field loops centered
at the midplane with alternating sense of circulation). The latter
yields two loops, one centered above the equator and the other
below, with the same sense of circulation. The $\sin(\log(r/h))$
modulation gives $\omega/\Omega_H\approx 0.44$, $\efluxem/\efluxma
\approx -3.6\%$, and $\tilde{L}=42\%$. The $|\sin(2\theta)|$
modulation gives $\omega/\Omega_H\approx 0.40$, $\efluxem/\efluxma
\approx -1.1\%$, and $\tilde{L}=8.3\%$. Increasing the number of
initial field loops therefore leads to a weak (factor of $2-3$)
decrease in $\efluxem/\efluxma$, while making the field strength
more uniform around the loop increases $\tilde{L}$ by a factor of
$2$ with a nearly constant $\efluxem/\efluxma$. Higher resolution
studies may better resolve these simulations and show weaker
dependence on field geometry.

We have also considered a purely vertical field geometry: $A_\phi
\propto r \sin\theta$.  In a Newtonian context this would
correspond to a uniform $z$ field in cylindrical coordinates.  The
field is normalized so that $\beta=p_{gas,max}/p_{mag,max} =100$
and $400$ in the equator of the torus. The outcome is different
from any of the other models.

The funnel field in the vertical field run is strong compared to
the disk field.  The accretion rate is larger, by a factor of $5$,
than the fiducial run.  In the early stages there is a brief net
outflow of energy from the black hole (although the total energy
released from the hole is negligible compared to the energy gained
at later times).  The $\beta = 100$ model has a high mean
efficiency; $\eflux/\mflux = 0.77$, compared to $0.82$ expected
for a thin disk. There is also a net {\it outflow} of angular
momentum from the black hole, with $\lflux/\mflux = -1.00$,
compared to $1.95$ expected for a thin disk.  The wind has a peak
asymptotic radial velocity $\tilde{v}^r = 0.94c$, attained near
the outer boundary, compared to $\tilde{v}^r=0.75c$ for the
fiducial run. Finally, the model has $\omega/\Omega_H\approx
0.41$, $\efluxem/\efluxma\approx -15\%$, and $\tilde{L}=79\%$. The
$\beta=400$ vertical field model has very similar properties,
which suggests that we are resolving the $\beta=100$ model.
Table~\ref{tbl3} summarizes measurements from the varying field
geometry models.

The models with net vertical field exhibit markedly different
behavior from the fiducial model.  It seems likely that some of
this difference is due to the axisymmetric nature of the model; in
3D matter can accrete between the vertical field lines without
having to push them into the hole.  That is, in 3D, it would be
easier for the hole to rid itself of the dipole moment that it
acquires in the net vertical field calculation.  But we cannot say
with any confidence what the outcome is until a full 3D experiment
on a disk with nonnegligible magnetic dipole moment.

\subsection{Numerical Parameters}\label{numparam}

\begin{deluxetable}{lllllll}
\tablewidth{0pt}

\tablecaption{Resolution Study\label{tbl4}}

\tablehead{

\colhead{Resolution}

& \colhead{$\efluxem/\efluxma$}

& \colhead{$\eflux/\mflux$}

& \colhead{$\lflux/\mflux$}

& \colhead{$\dot{a}/\mflux$}

& \colhead{$\mflux$}

& \colhead{$\tilde{L}$}

}

\startdata

 $64^2$        &-0.0528    & 0.914   & 1.630   & 0.036    & -0.159 & 0.55 \\
 $128\times64$ &-0.0438    & 0.841   & 1.420   & 0.121    & -0.165 & 0.23 \\
 $128^2$       &-0.0447    & 0.887   & 1.518   & 0.087    & -0.167 & 0.38 \\
 $256^2$       &-0.0316    & 0.874   & 1.274   & 0.198    & -0.186 & 0.27 \\
 $456^2$       &-0.0261    & 0.865   & 1.381   & 0.216    & -0.299 & 0.18 \\
\enddata
\tablecomments{ Numerator and denominators are separately time
averaged from $500<t<1000$ at the horizon.  This interval is chosen
so that all models are turbulent (in the lowest resolution model
turbulence decays shortly after $t = 1000$). The $456^2$ model is
the fiducial model. The nominal radiative efficiency is
$1-\eflux/\mflux$. A positive $\dot{a}/\mflux$ corresponds to a
spindown of the black hole because $\mflux < 0$. }

\end{deluxetable}

We have run the fiducial model at resolutions of $64^2$, $128^2$,
$128\times64$, $256^2$, and $456^2$. There is a weak dependence on
resolution in the sense that $\efluxem/\efluxma$ is smaller at
higher resolutions.  Lower resolution models do not sustain
turbulence for as long as high resolution models, so we average over
$500 < t < 1000$, when all models are turbulent. Table~\ref{tbl4}
gives a summary of results from the resolution study.  In every case
the nominal radiative efficiency is close to the thin disk value.

Resolution of the near-horizon region, where the energy density is
large, is also a concern, because our accretion rates are measured
there.  We have checked dependence on radial numerical resolution of the
near-horizon region by modifying the coordinate definition in equation
(7) to read $r = R_0 + e^{x_1}$ rather than $r = e^{x_1}$.  Increasing
$R_0$ from $0$ to the horizon radius increases the number of grid zones
located near the horizon.  We ran a model with $R_0 = 0.5$ and found no
significant difference from a comparable model with $R_0 = 0$.  This
suggests that we are adequately resolving the near-horizon region.

We also varied $R_{in}$ and $R_{out}$ and found no measurable
difference in $\efluxem$, $\efluxma$, $(B^r)^2$, and $\omega$ on
the horizon. We have moved $R_{out}$ from $40$ to $400$ and
$R_{in}$ from $0.7r_+$ to $0.98r_+$ and find negligible
differences in these quantities on the horizon.  The solution is
not sensitive to the location of the inner or outer boundary.
Moving the inner boundary of the computational domain outside the
horizon (e.g. $1.05r_+$) leads to strong reflections from the
boundary conditions and, ultimately, failure of the run. It is
possible that better inner boundary conditions or higher
resolutions could overcome this difficulty, but it seems cleaner
to simply leave the boundary inside the event horizon at $r=0.98
r_+$, out of causal contact with the rest of the simulation.

The model with larger $R_{out}=400$ does exhibit some new
features. The magnetic field lines in the funnel region have a
collimation factor of $10$ by the time they reach the outer
boundary.  At $R = 40$, however, both the $R_{out}=400$ model and
the fiducial model have a collimation factor of $5/2$.  By
$R_{out}=400$ the field lines are nearly cylindrical.  The peak of
the radial component of the asymptotic 3-velocity in the wind is
identical to the fiducial run with $\tilde{v}^r=0.75c$, indicating
little acceleration between $R = 40$ and $R = 400$.

The main numerical uncertainty in our experiments arise from the
floor on the density and internal energy.  We varied the floor
scaling from $\rho_{0,min} = 10^{-4}r^{-3/2}$ and $u_{min} =
10^{-6}r^{-5/2}$ to $\rho_{0,min} = 10^{-4}r^{-2.7}$ and $u_{min} =
10^{-6}r^{-3.7}$  (we chose these scalings so that $b^2/\rho_0$
would be nearly constant with radius in the funnel).  While this
significantly affects $b^2/\rho_0$, it does not otherwise affect
$\efluxem$ and $\efluxma$ or the mean values of $B^r$ and $\omega$
measured on the horizon.

We varied the floor normalization at $r = 1$ from the fiducial
values $(\rho_{0,min}, \IEDEN_{min}) = (10^{-4},10^{-6})$ to
$(10^{-5}, 10^{-7})$, and $(10^{-6} , 10^{-8})$.  This causes
almost no change in the flow near the horizon.  In the funnel,
however, we are at the limit of our ability to integrate the MHD
equations ($b^2/\rho_0\gg 1$). Our integration fails when we
attempt to use a mass density floor $\rho_{0,min}(r = 1) \lesssim
10^{-5}$ when $R_{out} \gg$ the outer edge of the initial torus.
Lower floors lead to faster outflows $\tilde{v}^r\gg 0.99c$ in the
funnel region, which are more likely to be numerically unstable.
These results hint that low density models will produce fast
outflows, but a confirmation awaits a more stable GRMHD algorithm.

The funnel region is difficult to integrate reliably, because when
$b^2/\rho_0 \gg 1$ small fractional errors in field evolution lead
to large fractional errors in the evolution of other flow
variables. This is a consequence of our conservative scheme, in
which all the dependent variables are coupled together by the
interconversion of primitive and conserved variables.  Evolution
of the MHD equations in nonconservative form (e.g. using an
internal, rather than total, energy equation), as in \cite{dh03},
may be slightly more robust, although De Villiers and Hawley
eventually experience similar problems in the funnel.  In any
event, the close correspondence between the numerical experiment
and the BZ model raises confidence in the results and suggests
that the magnetic field, if not the mass density and internal
energy density, is being evolved reliably.

\section{Discussion}\label{bzdiscussion}

We have used a general relativistic MHD code, HARM, to evolve a
weakly magnetized thick disk around a Kerr black hole.  Our main
result is that we find an outward electromagnetic energy flux on
the event horizon, as anticipated by \cite{bz77}.  The funnel
region near the polar axis of the black hole is consistent with
the Blandford-Znajek model.  The outward electromagnetic energy
flux is, however, overwhelmed by the inward flux of energy
associated with the rest-mass and internal energy of the accreting
plasma. This result essentially confirms work by
\citet{ga97,lop99} that suggested the BZ luminosity should be
small or comparable to the nominal accretion luminosity
($\tilde{L}\lesssim 1$).

One of our models discussed here, however, begins with a vertical
field threading the torus, exhibits a brief episode of outward
{\it net} energy flux.  This appears to be a transient associated
with the initial conditions.  The same model exhibits a steady
{\it net} outflow of angular momentum from the black hole.  Of all
our models, the vertical field model has the largest negative
$-\efluxem/\efluxma\approx 15\%$ (ratio of the electromagnetic
energy flux to ingoing matter energy flux) and largest
$\tilde{L}=\efluxem/(-\eps\mflux)\approx 80\%$ (ratio of
electromagnetic luminosity to nominal accretion luminosity). This
suggests that the BZ effect could play a significant role if the
disk has a net dipole moment and accumulates magnetic flux that
crosses the horizon.  This possibility will be considered in
future work.

Consistent with the results found earlier by \cite{dhk03}, we find
that our models can be divided into four regions: (1) a ``funnel''
region with $b^2/\rho_0 \gtrsim 1$ and $\beta \ll 1$; (2) a corona
with $1 \lesssim \beta \lesssim 3$; (3); an equatorial disk with
$\beta > 3$; and (4) a plunging region between the disk and event
horizon with $\beta \sim 1$ and a nearly laminar inflow from the
disk to the black hole.  We also find no feature in the surface
density at or near the ISCO (see figure~\ref{inflowcompplot}),
which agrees with the results by \citet{kh02,dhk03} and consistent
with \cite{rb97}. This is contrary to the sharp transition
predicted by thin disk models and used by XSPEC to fit X-ray
spectra.

We have shown that the funnel region is nearly force-free, and is
well described by the stationary force-free magnetosphere model of
\cite{bz77}, for which we have presented a self-contained
derivation in Kerr-Schild coordinates.  We find agreement between
the BZ model and our simulations in measurements of energy flux,
magnetic field, efficiency of accretion, and spindown power
output.  In all cases we find that in the force-free region the
field rotation frequency is about half the black hole spin
frequency, $\Omega_H \equiv a/(2 r_+)$. This spin frequency
maximizes electromagnetic energy output from the hole.  This
result is consistent with expectations of \cite{mt82} and the
force-free numerical results of \cite{kom01}.

We have also compared the time-average of the plunging region in
our fiducial model with the stationary MHD inflow model of
\cite{gam99}, which assumes that the flow matches a cold disk at
the ISCO.  The inflow model matches the simulated rest-mass flux
and electromagnetic flux of energy and angular momentum
surprisingly well, particularly considering the strongly variable
nature of the simulated flow in the plunging region. The inflow
model fails to match other aspects of the flow, such as the radial
component of the four-velocity.  This is mainly due to the finite
temperature of the simulated flow; the inflow solution assumes
zero temperature.  It is slightly surprising that the total
angular momentum flux is close to the value predicted by the zero
temperature inflow solution, and $20\%$ less than what is
predicted by the thin disk, yet the total energy flux is almost
exactly what is predicted by the thin disk.  It is as yet unclear
whether this is due to coincidence or conspiracy.

For a set of models similar to the fiducial model, the ratio of
electromagnetic to matter energy fluxes is sensitive to the black
hole spin, reaching $-7\%$ for $a \sim 1$. The evolution is
sensitive to the initial field geometry.  Models with a net
vertical field are more efficient, and more electromagnetically
active than models with comparable field strength but zero net
vertical field. Our models have a weak dependence on resolution in
the sense that as resolution increases the relative importance of
electromagnetic energy fluxes on the horizon diminishes.

With an $R_{out}=400$ model we demonstrate that an outgoing
electromagnetic energy flux can reach large radii.  The field in
the funnel region does not connect back into the disk.  Rather the
poloidal components lie parallel to the polar axis.  The field
lines are collimated by a factor of $5/2$ at $r = 40$ and by a
factor of $10$ at $r = 400$.  An outflow along the boundaries of
the funnel reaches a maximum $\tilde{v}^r\approx 0.75c$, but this
is sensitive to the value of our artificial density ``floor'': a
model with lower density reaches even larger radial velocities at
the outer boundary of the computational domain.

\citet{koide2002} have evolved a cold, highly magnetized uniform
density plasma ($\rho_0/p=0.06$, $b^2/\rho_0=10$) as it falls into
a rapidly spinning ($a=0.99995$) black hole in Boyer-Lindquist
coordinates for a time $\approx 14 G M/c^3$ using the MHD
approximation. This initial state does not correspond to an
accretion disk system. They demonstrated, however, that a
transient {\it net} energy extraction is possible from a spinning
black hole. Because of the short evolution time they are unable to
say whether the energy extraction process is possible in steady
state. \citet{koide2003} gives an expanded discussion of the above
system.

In contrast to the results of \citet{koide2002} and
\citet{koide2003}, we model a disk with an initially hydrodynamic
equilibrium fluid that is weakly magnetized.  We also use a
Kerr-Schild (horizon penetrating) coordinate system that avoids
potential problems associated with the treatment of inner boundary
condition in Boyer-Lindquist coordinates. In our simulation the
Balbus-Hawley instability drives turbulence and accretion in a
steady state where we evolve for a time $2000GM/c^3$.  We measure
a sustained outward electromagnetic energy flux that is smaller
than the inward matter energy flux (i.e. net {\it inward} energy
flux). Their model is evolved for too short a time to observe the
unbound mass outflow in the funnel region as seen by us and
\cite{dhk03,dh04}.

\citet{dh04}(hereafter DH) have also considered the numerical evolution
of weakly magnetized tori around rotating black holes.  Their models are
quite similar to ours in many respects, although they differ in that:
(1) their models are three dimensional while our models are
axisymmetric; (2) they use a nonconservative numerical method
\cite{dh03}; (3) DH use Boyer-Lindquist while we use Kerr-Schild
coordinates; (4) DH choose $\gamma = 5/3$ while we use $\gamma = 4/3$;
(5) DH's initial pressure maximum is located at $25 M$, while ours are
typically at $12 M$.  Our results for the energy and angular momentum
per baryon accreted from table~\ref{tbl2} can be compared to Table 1 of
DH by computing $\eflux/\mflux=\Delta E_i/\Delta M_i$ and
$\lflux/\mflux=\Delta L_i/\Delta M_i$. For models with $a = (0, 0.5,
0.9)$ DH find $\eflux/\mflux=(0.91,0.91,0.84)$ while we find
$\eflux/\mflux = (0.96,0.93,0.88)$.  For the same models DH find
$\lflux/\mflux=(3.1,2.6,1.9)$, while we find $\lflux/\mflux =
(3.1,2.6,1.7)$.  Given the differences in the models and numerical
methods, this quantitative agreement is remarkable.  Our models and De
Villier and Hawley's models also agree qualitatively in the sense that
both show a similar geometry of disk, corona, and funnel and both imply
that spin equilibrium is achieved at $a \sim 0.9$ (see \citealt{gsm2004}).

\cite{kom01} finds the BZ solution to be stable in force-free
electrodynamics, and \cite{kom02,kom04a} find the BZ solution to be
causal, but inconsistent with the membrane paradigm.  We find our
numerical solutions to be consistent with the BZ solution in the
low-density funnel region around the black hole. A numerical general
relativistic MHD study of strongly magnetized (monopole magnetic field)
accretion by \cite{kom04b} is also consistent with the BZ solution.  For
the strong field chosen he finds a considerably faster outflow (Lorentz
factors of $\approx 14$) than found in our models (Lorentz factors of
$\approx 1.5-3.0$).  Komissarov's model does not contain a disk.

The limitations of the numerical models presented here include the
assumption of axisymmetry and a nonradiative gas.  The effect of
axisymmetry can be tested by comparing our models with the three
dimensional models of \cite{dh04}; the angular momentum and energy per
accreted baryon in the two models differs by only a few percent.  In
addition the jet structure observed in \cite{dh04} is nearly
axisymmetric.  This is encouraging, although it is unlikely that an
axisymmetric calculation can capture the full range of possible
dynamical behavior in the accretion flow.

The radiation field, which we have completely neglected here, is likely
to play a significant role in the flow dynamics, through radiation force
on the outflowing plasma in the wind and through photon bubbles in the
disk \cite{gam98,socrates02}.  It will also, of course, play a
significant role in heating and cooling the plasma.  This is clearly
the most significant limitation of our calculation-- particularly from
the standpoint of comparison with observations-- and clearly the most
numerically difficult problem to overcome.

\acknowledgments

This work was supported in part by a NASA GSRP Fellowship Grant
S01-GSRP-044 to JCM, and NSF Grants AST-0093091 and PHY 02-05155.
Computations were done in part on {\tt platinum.ncsa.uiuc.edu}.  We
thank Stu Shapiro, Shinji Koide, Serguei Komissarov, Julian Krolik,
and John Hawley for comments.

\end{document}